\documentclass{agujournal2019}

\usepackage{etoolbox}
\makeatletter
\patchcmd\@combinedblfloats{\box\@outputbox}{\unvbox\@outputbox}{}{}
\makeatother

\usepackage{url} 
\usepackage{lineno}
\usepackage{ulem}
\usepackage{siunitx}

\draftfalse
\journalname{JGR: Space Physics}

\begin{document}

\title{20 Years of ACE Data: How Superposed Epoch Analyses Reveal
Generic Features in Interplanetary CME Profiles}

\authors{F. Regnault\affil{1}, M. Janvier\affil{1}, P. D\'emoulin\affil{2}, F. Auch\`ere\affil{1}, A. Strugarek\affil{3}, S. Dasso\affil{4,5}, and C. No\^us\affil{6}}

\affiliation{1}{Institut d’Astrophysique Spatiale, Universit\'e Paris-Saclay, CNRS, 91405 Orsay, France}
\affiliation{2}{LESIA, Observatoire de Paris, Universit\'e PSL, CNRS, Sorbonne Universit\'e, Univ. Paris Diderot, Sorbonne Paris Cit\'e, 5 place Jules Janssen, 92195 Meudon, France}
\affiliation{3}{D\'epartement d'Astrophysique-AIM, CEA/DRF/IRFU, CNRS/INSU, Universit\'e Paris-Saclay, Universit\'e Paris-Diderot, Universit\'e de Paris, F-91191 Gif-sur-Yvette, France}
\affiliation{4}{CONICET, Universidad de Buenos Aires, Instituto de Astronom\'\i a y F\'\i sica del Espacio, CC. 67, Suc. 28, 1428 Buenos Aires, Argentina}
\affiliation{5}{Universidad de Buenos Aires, Facultad de Ciencias Exactas y Naturales, 
Departamento de Ciencias de la Atm\'osfera y los Oc\'eanos and Departamento 
de F\'\i sica, 1428 Buenos Aires, Argentina}
\affiliation{6}{Laboratoire Cogitamus, 1 3/4 rue Descartes, 75005 Paris, France}


\correspondingauthor{Florian Regnault}{florian.regnault@u-psud.fr}

\begin{keypoints}
\item ICMEs relatively faster than the solar wind at 1 AU show signs of compression throughout their substructures.
\item ICME profiles with and without a detected magnetic cloud are similar, showing that all may have MCs but crossed at different locations.
\item There is a wider range of population of ICMEs emitted during the active phase of the solar cycle while the typical profiles are the same.
\end{keypoints}


\begin{abstract}

Interplanetary coronal mass ejections (ICMEs) are magnetic structures propagating from the Sun's corona to the interplanetary medium. 
With over 20 years of observations at the L1 libration point, ACE offers hundreds of ICMEs detected at different times during several solar cycles, and with different features such as the propagation speed.
We investigate a revisited catalog of more than 400 ICMEs using the superposed epoch method on the mean, median and the most probable values of the distribution of magnetic and plasma parameters. We also investigate the effects of the speed of ICMEs relative to the solar wind, the solar cycle, and the existence of a magnetic cloud on the generic ICME profile.
We find that fast-propagating ICMEs (relatively to the solar wind in front) still show signs of compression at 1~au, as seen by the compressed sheath and the asymmetric profile of the magnetic field. While the solar cycle evolution doesn't impact the generic features of ICMEs, there are more extreme events during the active part of the cycle, widening the distributions of all parameters. Finally, we find that ICMEs with or without a detected magnetic cloud show similar profiles, which confirms the hypothesis that ICMEs with no detected magnetic clouds are crossed further away from the flux rope core.
Such a study provides a generic understanding of processes that shape the overall features of ICMEs in the solar wind, and can be extended with future missions at different locations in the solar system.

\end{abstract}

\section*{Plain Language Summary}
Interplanetary coronal mass ejections (ICMEs) are magnetized clouds of solar material expelled from the Sun's atmosphere into the interplanetary medium. 
In the present article, we use plasma and magnetic data from the NASA ACE mission positioned at the Lagrange 1 point, and perform statistical studies over the 400 ICMEs detected over the last 20 years. Such an approach, called a superposed epoch analysis, provides a temporal description of the generic features of ICMEs.
In particular, we find that ICMEs that propagate faster than their surrounding solar wind still show compressed regions of plasmas and stronger magnetic field.
While we find that the 11-year solar cycle doesn't affect the generic properties of ICMEs, we observe more extreme events during the active phase.
Finally, we also find that a subset of ICMEs, called magnetic clouds and often associated with a stronger and more coherent magnetic field, have the same properties as other ICMEs. This confirms that most ICMEs may be magnetic clouds but which detection is limited by the spacecraft crossing.
Such a study, based on long-term heliospheric missions, allows us to probe physical processes that occur during the propagation of ICMEs, which can help better predict space weather and its consequences.

%
\section{Introduction}
\label{sect_intro}

Coronal mass ejections (CMEs) are the manifestation of violent events occurring in the atmosphere of the Sun \cite{Chen2011}. Detected in remote-sensing instruments such as coronagraphs or heliospheric imagers, these large structures, seen via Thompson scattering, are associated with plasma and magnetic field ejected from the Sun. The state of the corona prior to a CME ejection indicates the existence of a structured magnetic field, in particular the possible existence of a twisted magnetic field or flux rope \cite{Demoulin2008r}. Coronagraphic and heliospheric imaging can provide some clues on global dynamics of CMEs in their early propagation phase. For example, such data are used to infer the kinematics of events \cite{Sheeley1999,Lugaz2010, Davies2013}, with the multiplicity of observers providing constraints and accurate arrival times at planets \cite{Davis2009}. Such data can also be compared with simulations \cite<{\it e.g.,}>{Lugaz2011}, as well as be fitted by flux rope models to infer their geometrical properties \cite<such as in>{Thernisien2006,Wood2017}.

While CMEs are routinely detected with remote-sensing instruments, such data can be completed by direct measurements made by space probes when propagating away from the Sun into the heliosphere. They are then called interplanetary CMEs, or ICMEs. Compared with the ambient solar wind (SW), ICMEs are seen in time series of plasma and magnetic field parameters as a separate magnetic entity. The latter consists in a magnetic ejecta (ME) defined with a smooth and strong magnetic field \cite<{\it e.g.,}>{Winslow2015}. When plasma diagnostics are available, MEs are also associated with a low proton temperature $T_{\rm p}$\ (typically less than half of the expected SW temperature), and a low plasma $\beta$ which is the ratio of the thermal pressure over the magnetic pressure \cite<{\it e.g.,}>{Gosling1973, Richardson1995, Wang2005}. Due to different diagnostics available on different probes, several other criteria have also been given to define MEs \cite{Wimmer-Schweingruber2006, Zurbuchen2006}, such as: enhanced ion charge states \cite<{\it e.g.,}>{Fenimore1980}, enhanced helium abundance \cite<{\it e.g.,}>{Borrini1982} and counter-streaming suprathermal ($>$80 eV) electron beams \cite<{\it e.g.,}>{Gosling1987}.

For a large fraction of these MEs, a region of heated and compressed SW plasma, along with a highly varying magnetic field called the sheath is present at their front. This is due to SW material being accreted during the evolution of the ICME, as well as from the expansion of the CME \cite<{\it e.g.,}.>{Kaymaz2006}. Figure 17 in \citeA{Kilpua2017} illustrates the different SW properties in the sheath, where the magnetic field variations are considerably larger, and the temperature and density much higher than in the following ME. ICMEs that propagate in the SW with a velocity difference larger than the fast mode speed can also develop a fast forward shock.

Sometimes, the ME or part of it shows a smooth and large rotation of the magnetic field. Such cases are defined as magnetic clouds \cite<MCs,>{Burlaga1995,Burlaga1981} and are classically associated with the existence of a flux rope. The ratio of how many ICMEs show the existence of flux ropes is still under debate. At 1 au, many authors have reported that such configurations only occur in about one-third of ICMEs \cite<{\it e.g.,}>{Gosling1999,Bothmer1996,Cane2003,Huttunen2005,Wu2011}. However, multispacecraft observations such as in \citeA{Cane1997,Kilpua2011} have shown that the detection of flux rope is highly dependent on where the spacecraft crosses the structure. Then, many MEs are not detected as flux rope simply because the spacecraft crossing is too far from the axis or along the legs of the cloud, not allowing a certain detection of the twisted structure.

While interplanetary probes provide us with direct measurements of plasma and magnetic field parameters, compared with remote-sensing observations, these observations remain local. It is then difficult to obtain a global understanding of the structure of ICMEs. In some occurrences, multiple spacecraft crossings of the same event have been reported, for example with the STEREO missions \cite{Kaiser2008,Zhao2017} and that at the Lagrangian point 1 \cite<L1, {\it e.g.,}>{Ruffenach2012}, but these remain rare and are confined to case-studies. 

Other interplanetary probes like the Advanced Composition Explorer (ACE) \cite{Stone1998} have recorded plasma and magnetic field parameters for a continuous period and during a long interval of time ($>20$ years). As such, catalogs of ICMEs are available: some of them report the arrival dates of the different substructures of ICMEs (\citeA{Richardson2010}, hereafter referred to as the R\&C\ catalog, \citeA{Nieves2018} or the European-funded HELCATS project, \url{https://www.helcats-fp7.eu/index.html}), some provide information on the fit of MCs within these ICMEs \cite{Lynch2005, Lepping2010}, while some others focus on the normal of the shock of these ICMEs \cite{Feng2010, Wang2010}. These catalogs provide invaluable resources to look at a large sample of ICMEs.

Then, using statistical tools on these catalogs allows investigating generic features of ICMEs in details. In recent papers, \citeA{Janvier2014b, Janvier2015, Demoulin2016} investigated the generic shape of the MC axis as well as that of the shock at the forefront of ICMEs, by using lists of reported orientations of MC axis and shock normal. Such analyses can provide clues and constraints on what to expect in global numerical simulations of ICMEs.

Another technique is the so-called superposed epoch analysis (SEA, or Chree analysis, \citeA{Chree1914}), which allows one to superpose time series of parameters within well delimited structures (at least, bounded by one clear frontier such as a discontinuity). This technique has been used in a variety of cases, in particular in the SW for the investigation of corotating interaction regions \cite{Yermolaev2015}, ICMEs \cite{Janvier2019} and MCs \cite{Rodriguez2016, Masias-meza2016}. A recent study also investigated, using the SEA technique, the profile of the magnetic field twist within MCs \cite{LanabereEA20}.

\citeA{Masias-meza2016} analyzed how the generic profiles of MCs change depending on their propagation speed. Namely, the authors found that slowly propagating MCs tend to have a symmetric magnetic field profile, while fast-propagating MCs tend to have an asymmetric magnetic field profile, with a higher $B_{\rm tot}$ at the MC front compared with the rear. 
This study was recently extended by \citeA{Janvier2019} to catalogs of ICMEs detected by interplanetary probes at different heliospheric distances, with MESSENGER \cite{Solomon2007}, Venus Express \cite{Titov2006} and ACE. The results extended the conclusions found by \citeA{Masias-meza2016} to Venus and Mercury distances as well as to the broader class of MEs. 
  
As a follow-up to the two previous studies \cite{Masias-meza2016,Janvier2019}, we investigate in the following the generic profiles of ICMEs and their substructures for a large sample of ICMEs seen at 1~AU. 
The paper is organised as follows. In Section \ref{sect_dataset}, we present the set of data used throughout the study, and the additional work to prepare a thorough catalog of ICME frontiers. In Section \ref{sect_SEA}, we discuss how the SEA technique can be applied to this large set of ICMEs. 
In particular, we introduce the most probable value of the distribution, in complement of the mean and median that are typically used in the litterature, to better characterize the ICME properties. In Section \ref{sect_SE_cat_1}, we present SEA applied to categories of ICMEs classified in function of other parameters ({\it e.g.,}\  speed, solar cycle, presence of a MC), before discussing and concluding on the results in Section \ref{sect_conclusion}.

%

%
\section{Description of the Dataset}
\label{sect_dataset}

\subsection{ACE Data}
\label{sect_dataset_ACE}

The plasma and magnetic field parameters used in the present study come from two instruments on board of the Advanced Composition Explorer (ACE) satellite. The mission, with a stable orbit around the L1 libration point, has been recording continuously the SW since its launch on 25 August 1997.

The magnetic field measurement comes from the MAG instrument \cite{Smith1998}, which measures the local magnetic field intensity with a time resolution of 1~second. The other parameters come from SWEPAM (Solar Wind Electron, 
Proton and Alpha Monitor, \citeA{McComas1998}), which measures plasma parameters (such as the proton density $n_{\rm p}$, temperature $T_{\rm p}$\ and velocity $V_{\rm p}$) and the composition of the SW with a time resolution of 64 seconds. We use a 16-seconds time resolution (average of 16 measurements done in 16 seconds) for MAG data and 64-seconds for SWEPAM. All data are directly available online. In the following, we use the first data available, starting from late 1997, up to the end of 2017. For the MAG data, the quality of each point is flagged by 0, 1 or 2~: 0 means that the data measurements are readily exploitable, 1 means that the measurements have been made during a period of high nutation of the spacecraft ($\approx$ 4 hours) or during a spacecraft manœuver, and 2 means bad or missing data. For this study, we decide to take only the quality 0 data points. While for the SWEPAM data a value of -9999.9 indicates bad or missing data.

\subsection{Catalog of Events} 
\label{sect_catalogue}

The present study uses a list based on the Richardson and Cane (R\&C) list of ICMEs  \cite{Cane2003, Richardson2010}. This catalog lists ICMEs that were visually detected between the period of 1996 and 2019 (as of March 2020) and provides the dates of detection of the ICME disturbance (typically defined as the associated geomagnetic storm sudden commencement detected at Earth), as well as the start and end dates of each ICME.

R\&C\ used different data sets from different spacecraft to create the list ({\it e.g.,}\ ACE, WIND \cite{Harten1995}, SOHO \cite{Fleck2015}, IMP 8 \cite{Paularena1999}), depending on the availability of the data. The reported boundaries for each individual event is given with an hour accuracy.  
Furthermore, the delimitation of an ICME boundary can be a matter of debate, and it is difficult to locate unambiguously in fine scale observations : different authors use different small scale features or combination of features to mark the ICME boundaries ({\it e.g.,}\  a particular discontinuity, or the onset of bidirectional electrons, or a plasma composition signature, see also discussions, {\it e.g.,}\ in \citeA{Richardson2010}).
Therefore, the R\&C\ list serves as a guideline to give an initial idea of where the boundary may be in the ACE dataset. Then, we revisit this list for each individual event, looking at all the magnetic field and plasma parameters as reported below. We also use the plasma $\beta$ parameter as derived in the OMNI database (\url{https://omniweb.gsfc.nasa.gov/ftpbrowser/magnetopause/Reference.html#Details}). It assumes that the temperature of the alpha particles is given by $T_{\alpha} = 4 T_{\rm p}$ with $T_{\rm p}$\  the temperature of the protons, and that the electron temperature $T_{\rm e}$ is constant with $T_{\rm e} = 1.4\times 10^5 K$. Finally we use the $rmsBoB$, which is defined by \citeA{Masias-meza2016} as :
  \begin{eqnarray}
  rmsB(t) = \sqrt{\sum_{i=1}^3 \langle (B_i - \langle B_i \rangle)^2 \rangle} \label{eq_rmsB}\\
  rmsBoB(t) = \frac{rmsB(t)}{B(t)} \, ,     \label{eq_rmsBoB}
  \end{eqnarray}
where $\langle\rangle$ stands for the temporal average over short time intervals (64 seconds). $rmsBoB$ (t) quantifies the fluctuations of the magnetic field intensity, $rmsB(t)$, relative to the magnetic field intensity ($rmsB$ over $B$).

To define the boundaries of the sheath, which is a region of compressed and heated SW plasma ahead of the ME, we look for an increased density and temperature in this region with an enhanced magnetic field. Furthermore, the presence of a discontinuity, being most of the time a shock, ahead of the sheath allows us to define unambiguously the beginning of this region.

Within a ME, we expect an enhanced magnetic field and a lower proton temperature. The transition from the sheath to the ME is then marked by a decrease in temperature. We use $T_p < T_{\mathrm{exp}}/2$ as a guideline to find the ME boundaries, with $T_{\mathrm{exp}}$ being the temperature expected in the typical SW with the same speed \cite{Richardson1995,Demoulin2009d}. $\beta$ and $rmsBoB$\ are also expected to well delimit the time interval corresponding to the presence of a ME: a low $\beta$ ($< 1$) is expected in the ME since we have a low temperature (small thermal pressure) and an enhanced magnetic field (high magnetic pressure), while we expect low magnetic intensity fluctuations inside the ME (i.e low $rmsBoB$) due to the presence of an intense and a coherent magnetic structure.

Finally the end of the ME is the most difficult frontier to define. As the parameters are expected to return to their pre-ICME values, we look for an increase in temperature, and a decrease in magnetic field intensity. However, this transition is hard to define because these changes are typically smooth, long and progressive, without a sharp discontinuity. Over all the parameters we investigated, none appeared to be really better than the others to define the end of the ME. 

The difficulty in defining these frontiers can be due to different reasons: a ME can in some cases, but not always, be composed of a coherent magnetic field structure ({\it e.g.,}\  a flux rope), itself surrounded by magnetic flux from overlying coronal arcades dragged along when the flux rope was first ejected in the corona. Then, the plasma and magnetic field surrounding a coherent magnetic core structure may still be magnetically distinct from the SW and the sheath, but with properties of coronal plasma embedded in coronal field lines dragged along the propagating ME. Furthermore, magnetic reconnection occurring at different locations along the ME, either at its front or rear boundary \cite{Ruffenach2015}, can smooth out the transition between the different substructures. 
\citeA{Rodriguez2016} found signatures indicating that the flux rope and SW plasma can be mixed at the rear of the ME because of reconnection processes.

Note that while some frontiers can be the subject of discussion between different authors, it was shown in \citeA{Janvier2019} that since the SEA averages small changes occurring around boundaries that are roughly defined of the order of a few hours,
the resulting superposed epochs are not highly sensitive to the definition of these boundaries. Then, the data set for the revisited catalog for 498 ACE detected ICMEs has been put online (see the link at the end of the paper), where we provide the time of the front discontinuity, as well as the start and end time of the ME, the presence or not of a sheath, and finally the quality of each event. We define 4 qualities for each event: Quality 1 (123 events) corresponds to well define events, with unambiguously defined boundaries with all parameters, Quality 2 (201 events) corresponds to ICME events with clear substructures but for which boundaries can be sometimes difficult to define ({\it e.g.,}\ no clear discontinuity and smooth transitions between substructures) or missing data in some parameters, Quality 3 (101 events) corresponds to ICMEs with clear substructures but with large data gaps over some parameters, and finally Quality 4 (73 events) are ICMEs for which there are too large data gaps or the presence of an ICME was unclear. None of the Quality 4 ICMEs are considered in the following, so that the total sample in the following corresponds to 425 ICMEs.

\begin{figure}[t!]
\centering
\includegraphics[width = 14 cm]{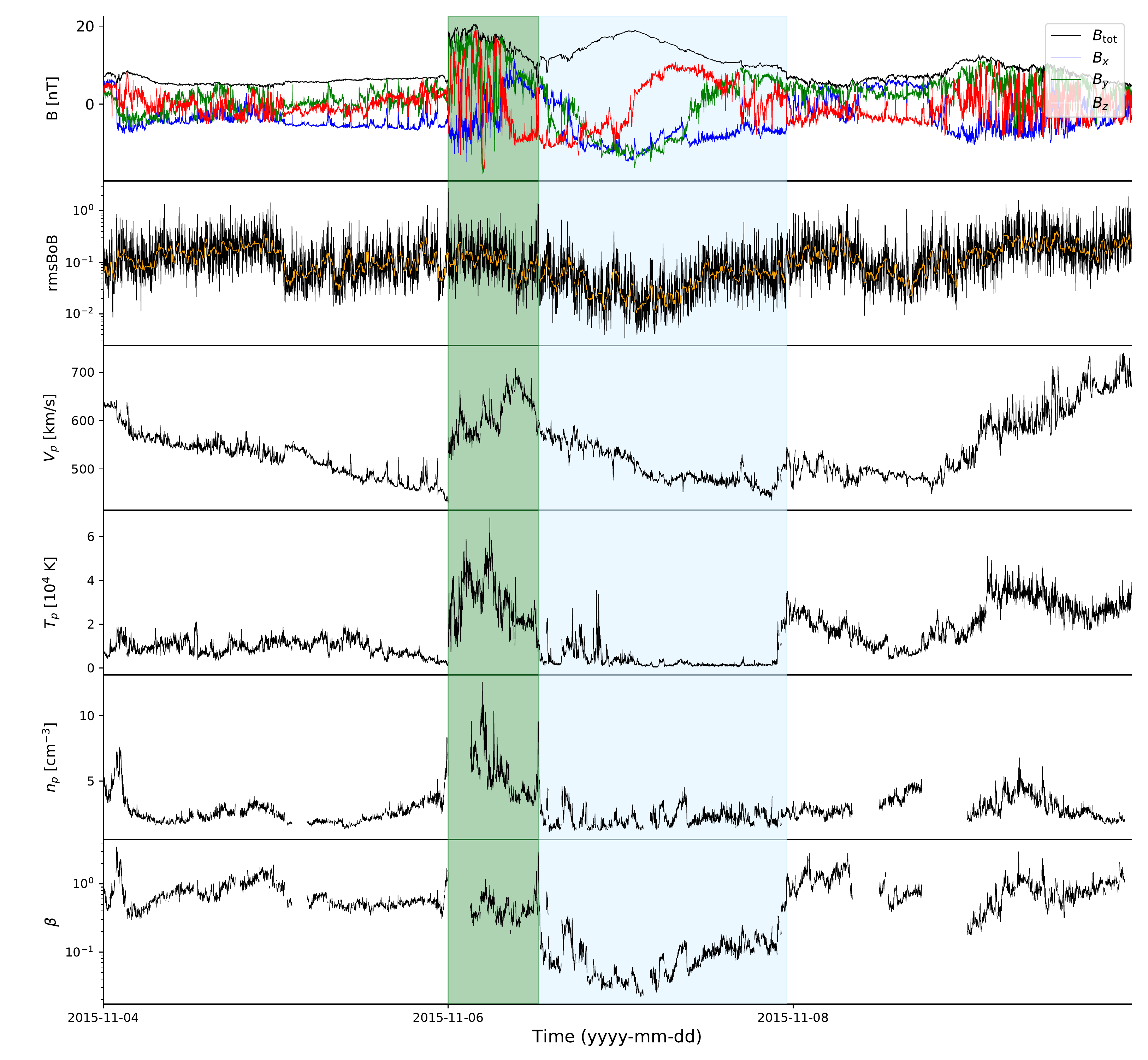}
\caption{Example of one ICME event observed by the ACE spacecraft. From top to bottom: time evolution of the magnetic field intensity $B_{\rm tot}$\ and its components in GSE coordinates, root mean square value of B over B $rmsBoB$\ (the yellow curve is the $rmsBoB$\ smoothed with a median filter computed using 25 points), bulk velocity $V_{\rm p}$, proton temperature $T_{\rm p}$, proton density $n_{\rm p}$\ and plasma parameter $\beta$. The sheath region is indicated by a green area and the ME with a blue area.}
\label{fig_ex_event}
\end{figure}

\subsection{Example of ICME from the ACE dataset} 
\label{sect_example}

Figure \ref{fig_ex_event} presents an example of an ICME seen within the ACE data. The top row shows the 
total magnetic field intensity $B_{\rm tot} = \sqrt{B_{\rm x}^2 + B_{\rm y}^2 + B_{\rm z}^2}$, as well as the $x$, $y$ and $z$ components in GSE (Geocentric Solar Ecliptic) coordinates. The $x$-direction is towards the Sun, $z$ is perpendicular to the ecliptic plane and $y$ completes the orthonormal vector basis. 
The second row shows the normalized magnetic fluctuations ($rmsBoB$) as defined in Equation \ref{eq_rmsBoB}.
The third to fifth rows show the velocity $V_{\rm p}$, the temperature $T_{\rm p}$\ and the density $n_{\rm p}$\ for the protons as measured by the SWEPAM instrument. 
The bottom row shows the plasma beta parameter $\beta$ as defined in the OMNI database. 

 Some typical characteristics of a sheath and a ME, respectively highlighted by the light green and blue areas, are present in Figure \ref{fig_ex_event}, as described in Section \ref{sect_catalogue}. 
The regions before and after the ICME are referred below as the pre-ICME and wake regions, respectively.
A sudden increase of $B_{\rm tot}$, $V_{\rm p}$, $T_{\rm p}$, $n_{\rm p}$\ is present at the beginning of the sheath.
The plasma $\beta$ remains almost constant within the sheath. 
The end of the sheath and the transition to the ME is highlighted by a decrease in fluctuations of $B_{\rm tot}$, in the proton temperature, as well as in $\beta$. The proton speed decreases inside the ME all the way to its rear boundary.
While the density increases within the sheath, inside the ME its values are similar to that of the SW.
The end of the ME is highlighted by an increase of $rmsBoB$, temperature and plasma $\beta$.

%
\section{Superposed Epoch Analysis Applied to ICME Events} 
\label{sect_SEA}

\subsection{Method}
\label{sect_SEA_Method}

In the following, we present the SEA applied to a large number of ICMEs with time series of the different {\it in situ}\ physical parameters. Doing so allows us to investigate whether typical profiles exist for each physical parameter measured by ACE and what these are. 

First, since not all ICME events have the same duration, we normalize all of them in time. However, since the sheath and the ME do not necessarily have the same relative duration, a typical ICME profile needs to be represented by a typical ratio of time interval between the sheath and the ME. The duration ratio between the ME and the sheath was defined to be 3:1 in \citeA{Masias-meza2016}. 
Our catalog has 330 ICME events of Quality $\leq 3$ with a well defined sheath. This multiplies by a factor 7.5 the number of events from previous studies. In our sample, we checked the distribution of the ME to sheath duration ratio over all the sample of events and we found that the distribution is asymmetric with a positive skewness (more probable values for lower ratio).
This means that the mean, median and most probable ratios are not the same. However, the most probable value, mean and median values all scale between 1.5 and 3.5.
Then, we still decide to define the ratio as 3:1 so as to compare with the SEA results found by \citeA{Masias-meza2016}, knowing that another ratio would give the same results, but on a different normalized scale. 
Then, all the sheath duration are rescaled to the normalized length of 1 and the ME duration are normalized to 3. This choice is only a matter of visualization and will no affect the results of our study.

Second, in order to compute the profiles of ICMEs including its surrounding SW, we define two other time intervals: the pre-ICME SW (hereafter, pre-SW) and the post-ICME SW (hereafter, the wake). The time interval chosen to represent both is as long 
as the ME for each event. 
Then, we bin all parameters in time for each ICME, which results in the same number of points in time for all events. We take 50 bins for the sheath and 150 bins for the ME. For the pre-SW, we take 1 ME duration before the beginning of the sheath and the same duration after the end of the ME for the wake which lead to 150 bins for both of these areas.

Usually, the SEA results are summarized by either the mean of the distribution, or its median in each time bin 
\cite{Yermolaev2015, Rodriguez2016, Masias-meza2016, Janvier2019}. 
However, any chosen single value is a reduction of a distribution and does not capture all of its characteristics. For example, for a normal (and thus symmetric) distribution, the mean, the median and the most probable value are all equal, however the distribution spread is another independent parameter of the distribution.
 Furthermore, for more general distributions, other characteristics such as their spread or their symmetry (or lack thereof) are not represented by any of these single values.

In the case of MEs, a log-normal distribution of plasma and magnetic field parameters was observed  by \citeA{Mitsakou2014} and \citeA{Rodriguez2016}. Hence, throughout our study, we represent SEA results either by means of stacked histograms (therefore showing the whole distribution for each time bin), or a combination of the median, mean and the most probable value. This latter value is computed by using the parameters of the observed log-normal distribution, which are obtained thanks to a maximum likelihood estimation algorithm (for an example of a typical asymmetric distribution found for the parameters in one time bin, and the method used to find the mode, see \ref{app_lognorm}). Such a combination of mean, median and most probable value, in the case of a non-symmetrical distribution, can give a better knowledge of the distribution shape as opposed to when providing only a single value. Finally, we notice that the asymmetry of the distribution of vector quantities such as $B$ is somewhat expected, as follows. If the components of the magnetic field (or any other vector quantity) were to follow a Gaussian distribution, then by construction, the associated magnitude, $B$, would follow a degree 3 Rayleigh distribution, that is an asymmetric distribution with a positive skewness resembling a log-normal. A careful look at the vector distributions for the present parameters shows that they are not exactly a degree 3 Rayleigh distribution, and a deeper study on the origins of such a log-normal like distribution is left for further studies.

In the downloaded ACE data, some data points are flagged as "bad or missing data", so that the 20 years data set has some data gaps. Unfortunately, some data gaps are within some events and last long enough to spoil the entire event ({\it e.g.,}\  by hiding a transition between two sub-structures). As we divide all ICMEs in two parts (pre-SW + sheath and ME + wake), we keep the entirety of one part if both pre-SW + sheath (respectively ME + wake) have enough data, i.e. when there are less than 30\% of data gaps. The SEA is then applied separately for these two parts, so that for a given superposed epoch, there can be a different number of events between the pre-SW + sheath part and ME + wake part.

\begin{figure}[t!]
\centering
\includegraphics[width = 14 cm]{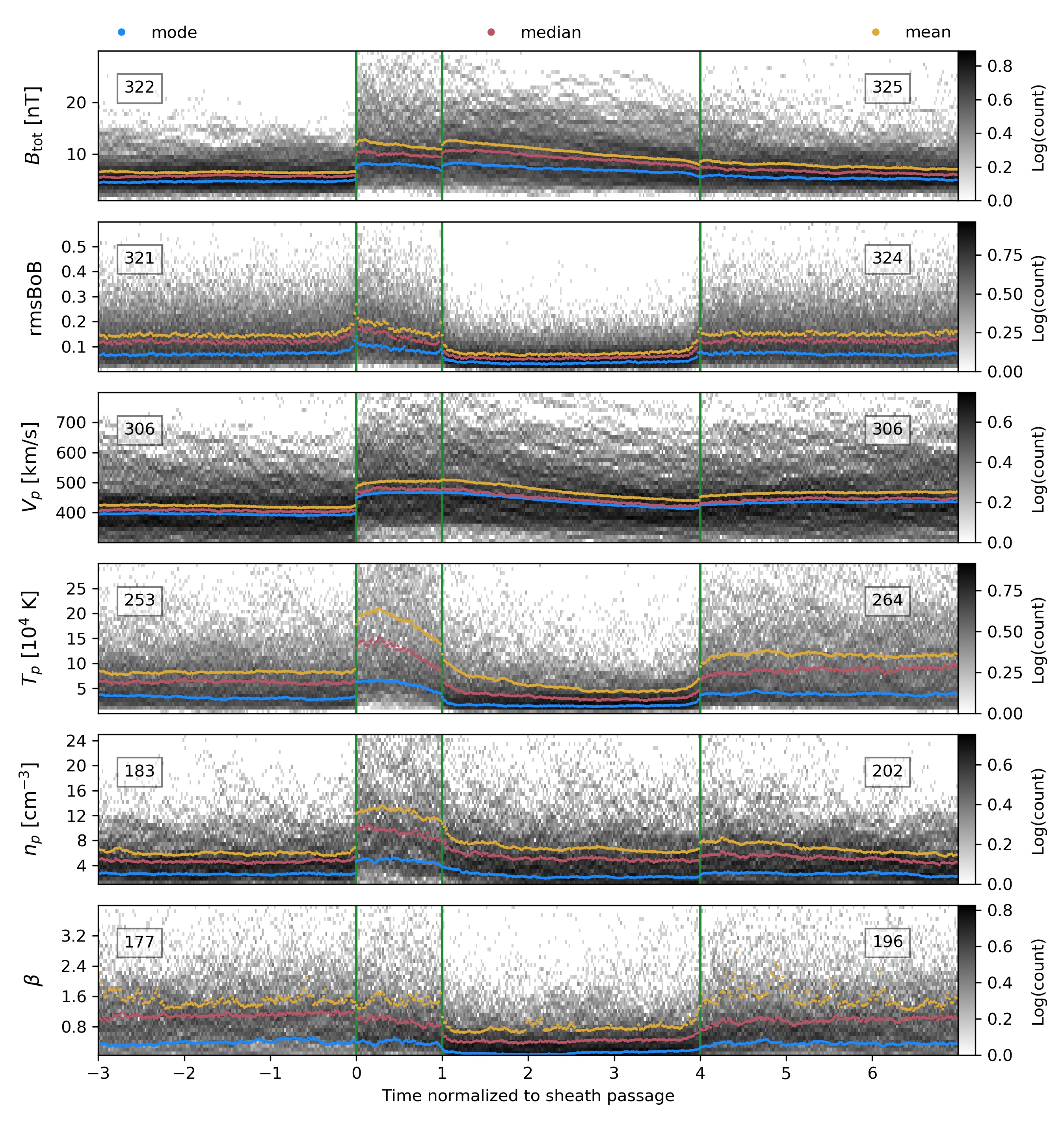}
\caption{ 
Superposed epoch analysis of all ICMEs events with a sheath from the revisited R\&C list. In the background, the color scale shows the frequency of each physical parameter. The frontiers between the pre-SW, the sheath, the ME and the wake are shown by the green vertical lines. 
Each region (pre-SW + sheath and ME + wake) has a different number of events depending on the availability of the data for each physical parameter.  This number is written on the left for the first part (pre-SW + sheath) and the right for the second part (ME + wake). 
From top to bottom: Magnetic field intensity $B_{\rm tot}$, root mean square value of B over B, noted $rmsBoB$, bulk velocity $V_{\rm p}$, proton temperature $T_{\rm p}$, proton density $n_{\rm p}$, plasma parameter $\beta$.  Yellow curves represents the mean, magenta ones the median and blue the most probable value obtained from the fit of the data by a log normal distribution in each time bin. The time is set such that $t = 0$ corresponds to the beginning of the sheath and $t = 1$ corresponds to its end. }
\label{fig_SE_sheath}
\end{figure}

\subsection{SEA on ICMEs with a Sheath}
\label{sect_SE_sheath}

In the following, we apply the SEA to all events with a sheath in the revisited R\&C\ list.
Figure \ref{fig_SE_sheath} presents the SEA as a stacked histogram, on the same parameters as in Figure \ref{fig_ex_event}.
For each time bin of the superposed epoch on one parameter, the logarithm of the number of counts is represented with grey levels. Dark areas correspond to more frequent values, while the lighter ones correspond to less frequent values.  Such a representation provides a better view  of the distributions and their evolution in time, and gives a straightforward understanding of the most probable value, the median and the mean of the distribution, each represented by the blue, magenta and yellow lines respectively.

\begin{table}
\begin{center}
\begin{tabular}{lcccc}
\hline
 & \multicolumn{1}{c}{pre-SW} &\multicolumn{1}{c}{sheath} &\multicolumn{1}{c}{ME} &\multicolumn{1}{c}{post SW} \\
\hline
$B_{\rm tot}$ [nT] & 4.7 $\pm$ 0.3 & 7.9 $\pm$ 0.9 & 7.2 $\pm$ 1.8 & 5.4 $\pm$ 0.7 \\
rmsBoB [$\times 10^{-2}$] & 7 $\pm$ 2 & 9 $\pm$ 4 & 4 $\pm$ 2 & 7 $\pm$ 1 \\
$V_p$ [km/s] & 395 $\pm$ 7 $\,$ $\,$ & 465 $\pm$ 12 $\,$ & 437 $\pm$ 52 $\,$ & 434 $\pm$ 9 $\,$ $\,$ \\
$T_p$ [$\times 10^4$ K] & 3.2 $\pm$ 0.8 & 5.9 $\pm$ 2.4 & 1.6 $\pm$ 0.8 & 3.9 $\pm$ 0.6 \\
$n_p$ [cm${}^{-3}$] & 2.7 $\pm$ 0.3 & 4.8 $\pm$ 0.8 & 2.4 $\pm$ 1.2 & 2.7 $\pm$ 0.6 \\
$\beta$ & 0.4 $\pm$ 0.2 & 0.4 $\pm$ 0.1 & 0.1 $\pm$ 0.1 & 0.4 $\pm$ 0.1 \\
\hline
\end{tabular}
\caption{Most probable values from the distribution of the magnetic and plasma parameters, averaged over the time period of each substructure (pre-SW, sheath, ME, and wake) for all ICMEs with a sheath (see Figure \ref{fig_SE_sheath}). Three times the standard deviation is included after $\pm$.}
\label{tab_all_event_wo_NS}
\end{center}
\end{table}

For all the graphs, the mean is always greater than the median, which is itself always greater than the most probable value (or the mode) of the log-normal distribution fitted to the distribution of each superposed epoch bin.
This is due to the asymmetry of the distributions which are typically well approximated by log-normal distributions, as was shown in the study of \citeA{Rodriguez2016} on MCs.

Here we generalize the results to ICMEs.  While the mean tends to be skewed by extreme events, the median and the most probable value are less sensitive to these events.
Therefore, in the following, we will describe the profiles using the most probable value as a proxy to the typical behavior of ICMEs, while the relative variation of the median and mean values give us an understanding of how the distribution evolves across ICMEs. Globally, the three quantities, mean, median and most probable value, mostly follow each other closely, which reflects the change in the width of the distribution (Figure \ref{fig_SE_sheath}).
In Table \ref{tab_all_event_wo_NS}, we also indicate the most probable value in the pre-SW, sheath, ME and wake of the SEA for all events with a sheath. 
The values correspond to an average of the most probable value across the time interval corresponding to each substructure.
We also add three times the standard deviation, which includes both the fluctuations and the global temporal trend present in each region.

From the profiles in Figure \ref{fig_SE_sheath}, we see a clear discontinuity of all parameters between the pre-SW and the sheath (except for $\beta$), with an enhancement of these parameters in the sheath. 
In order to quantify the enhancements from the pre-SW to the sheath, we define the ratio $r_{x} = \langle x_{sheath}\rangle / \langle x_{pre-SW}\rangle$, with $\langle \rangle$ being the temporal average of the most probable value and $x$ the parameter. This ratio is equal to 1.7 $\pm$ 0.2 for $B_{\rm tot}$, 1.8 $\pm$ 0.9 for $T_{\rm p}$\ and 1.8 $\pm$ 0.4 for $n_{\rm p}$.
Despite these physical changes, these similar $r_{x}$ values imply that the plasma $\beta$ remains almost constant. We also observe a peak of the most probable value of the $rmsBoB$\ at the frontier between the pre-SW and the sheath due to the abrupt change of the magnetic field intensity there (Figure \ref{fig_SE_sheath}, second panel). 
  
The ME front is marked by a second abrupt transition in the magnetic field intensity. 
The magnetic field profile of the ME is therefore asymmetric, with a higher
intensity at the front. This was already pointed out in other ICME
statistical studies, {\it e.g.,} in the study of \citeA{owens2005}.
The speed decreases monotonically within the ME, with a constant slope. This is generally
interpreted as an effect of the expansion of the ME while it propagates \cite<>[and references therein]{Demoulin2008,Gulisano2010}.  This expansion is induced by the decrease with solar distance of the total SW pressure, so that the pressure in the ME should also decrease \cite{Demoulin2009}.  
The temperature decreases by a ratio of $r_{T_{\rm p}} = 3.7 \pm 2.4$ from the sheath to the ME and the density decreases by a ratio of $r_{n_{\rm p}} = 2.0 \pm 1.1$ going to values similar to those of the pre-SW plasma.

The boundary between the ME and the wake is typically the most difficult to define out of the three boundaries (Section \ref{sect_catalogue}). However, as we also observe sharp transitions for $rmsBoB$\ and $T_{\rm p}$\ parameters at the rear edge of the ME, the SEA shows that this frontier is overall well defined.

The parameters $rmsBoB$\ and $\beta$ are good indicators of the ME extension because they have a sharp decrease and an increase in their values at the beginning and at the end of the ME, respectively, while they remain higher and almost constant within the sheath and the wake (we still notice that $rmsBoB$\ has a nearly uniform decrease within the sheath away from the shock). The decrease of the $rmsBoB$\ value in the ME is due on one hand to the crossing of a coherent magnetic structure and on the other hand to the increase of the magnetic field inside the ME, see Equations \ref{eq_rmsB} and \ref{eq_rmsBoB}.
The decrease of $\beta$ combines the decreases of $T_{\rm p}$\ and $n_{\rm p}$\ with an enhanced magnetic field inside the ME.

In the wake of the ICME, 
mainly $B_{\rm tot}$\ and $V_{\rm p}$\ do not go back to their pre-SW values straight away, this being clearer in the tail of the ICME distribution, as shown with the mean (Figure \ref{fig_SE_sheath}). 
We observe an increase by a factor 1.10 $\pm$ 0.03 of wake speed compared with the pre-SW speed. Then, we conclude that the SW conditions, in particular the speed, take at least a time interval as long as a full ME length to recover to its original state. The ICME disturbance is therefore felt for a much long period than just the ICME interval length, as was already pointed out in the studies of \citeA{Temmer2017}.

\begin{figure}[t!]
\centering
\includegraphics[width = 12 cm]{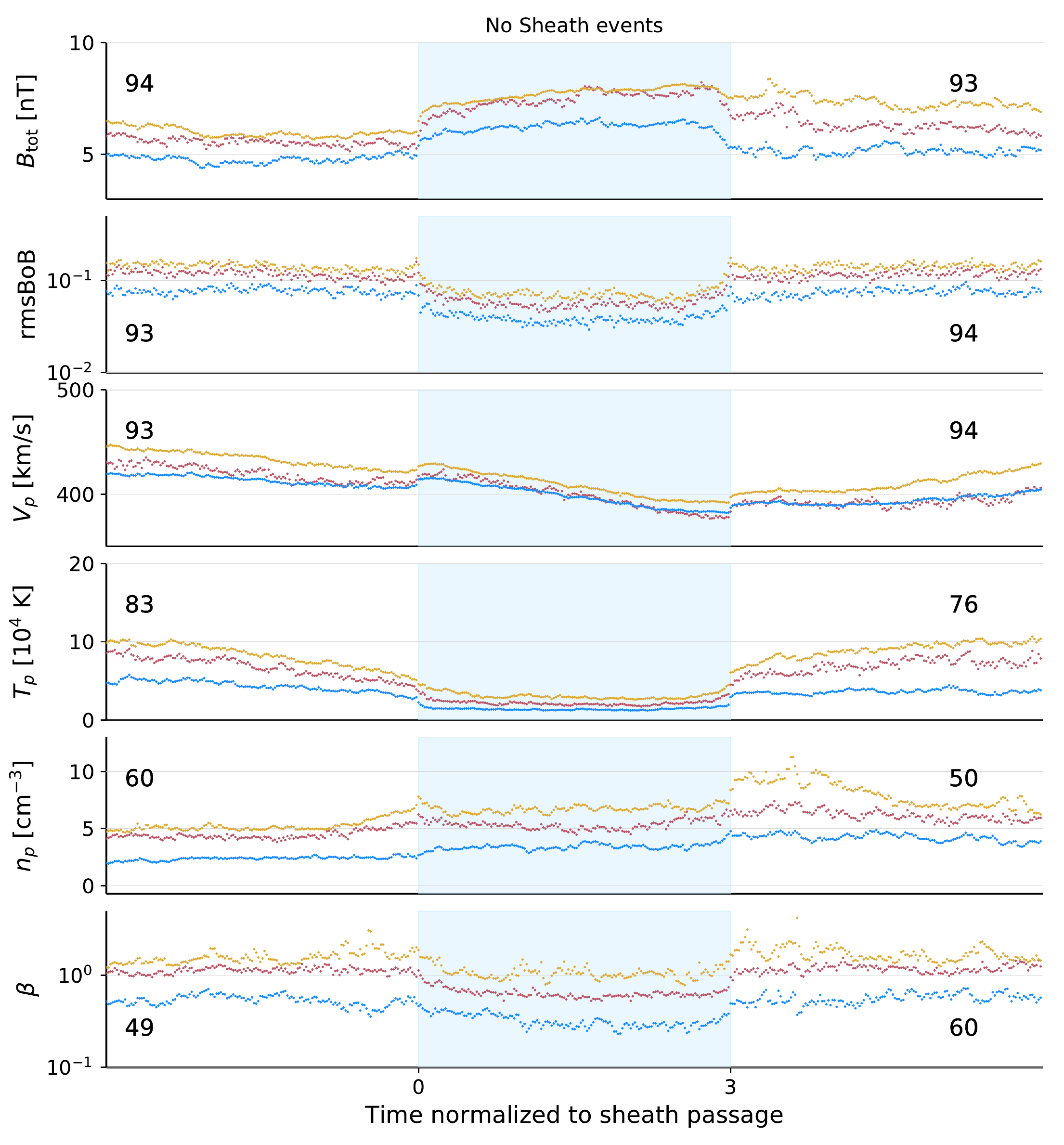}
\caption{Superposed epoch analysis on all ICMEs without a visible
sheath. The physical parameters shown are the same as in Figure \ref{fig_SE_sheath}. 
The blue area indicates the ME. The yellow dots represent the mean values of the parameter, the magenta ones the median values, and the blue ones the most probable values. The numbers on the top left (right) of each graphs are the number of events with more than 70\% of available data for the pre-SW (ME and the wake).}
\label{fig_SE_NS} 
\end{figure}

\begin{table}
\begin{center}
\begin{tabular}{lccc}
\hline
 & \multicolumn{1}{c}{pre-SW} &\multicolumn{1}{c}{ME} &\multicolumn{1}{c}{post-SW} \\
\hline
$B_{\rm tot}$ [nT] & 4.8 $\pm$ 0.5 & 6.2 $\pm$ 0.7 & 5.1 $\pm$ 0.5 \\
rmsBoB [$\times 10^{-2}$] & 8 $\pm$ 2 & 4 $\pm$ 2 & 7 $\pm$ 2 \\
$V_p$ [km/s] & 413 $\pm$ 15 $\,$ & 398 $\pm$ 33 $\,$ & 394 $\pm$ 13 $\,$ \\
$T_p$ [$\times 10^4$ K] & 4.3 $\pm$ 2.1 & 1.4 $\pm$ 0.5 & 3.6 $\pm$ 0.8 \\
$n_p$ [cm${}^{-3}$] & 2.4 $\pm$ 0.5 & 3.4 $\pm$ 0.8 & 4.2 $\pm$ 1.0 \\
$\beta$ & 0.5 $\pm$ 0.2 & 0.3 $\pm$ 0.2 & 0.6 $\pm$ 0.2 \\
\hline
\end{tabular}
\caption{Most probable values from the distribution of the magnetic and plasma parameters, averaged over the time period of each substructure (pre-SW, ME, and wake), for all events without a sheath (see Figure \ref{fig_SE_NS}). Three times the standard deviation is included after $\pm$. }
\label{tab_NS}
\end{center}
\end{table}

\subsection{SEA on ICMEs without a Sheath}
\label{sect_SE_NS}

We next investigate the 95 events in our list that do not have a clear sheath at the forefront of the ME. These events are then superposed following the SEA technique, and the results are reported in Figure \ref{fig_SE_NS}. The parameters shown are the same as those shown in Figure \ref{fig_SE_sheath}. In these graphs, as well as all the following ones, we only show the mean, median and most probable value as proxies of the shape for the distribution of each time bins, and not the histogram in the background, so as to not overload the graphs.

The time average of the most probable value for the same parameters as in Table \ref{tab_all_event_wo_NS} are reported in Table \ref{tab_NS} for events without a sheath.
The large standard deviation of the temperature is due to the decreasing profile of the temperature near the beginning of the ME. Taking into account the parameters within three standard deviations, the events with and without sheath propagate in a comparable SW.

The expansion of the ME causes a higher front speed than the rear. This induces a large standard deviation for the speed averaged over the ME duration. However, if we compare the front and rear speed of events with and without a sheath, we find that ME associated with a sheath propagate faster than those with none: indeed, for events without a sheath, the front speed is 410 km/s  and 390 km/s for the rear, while the front speed is 470 km/s and the rear speed 420 km/s for events with a sheath (Figures \ref{fig_SE_sheath} and  \ref{fig_SE_NS}).\citeA{Chi2016} observed a difference in speed between ICME with and without shocks. We can thus interpret events without a sheath as events that did not propagate fast enough to accumulate enough SW material at their fronts (within the limits of having only data at a given solar distance, so ignoring the history of the solar wind interaction).

Another important difference is that the magnetic field profile is more symmetric for MEs without a sheath compared to those that have a sheath at their front. An interpretation was given in \citeA{Masias-meza2016} and \citeA{Janvier2019}, that a more symmetric profile results from a quasi-equilibrium between the conditions encountered in the surroundings of the ME (pre-ICME and post-ICME SW): without the extra sheath, the total pressure in the front and with the low $\beta$ condition present within an ME, the magnetic profile within MEs without a sheath is expected to be more symmetric.

Other ME characteristics are similar with and without a sheath.
The $rmsBoB$\ has a sharp decrease at the beginning up to the end of the ME. The $rmsBoB$\ decreases by a factor 2.0 $\pm$ 1.1 inside the ME while the expected decrease of plasma $\beta$ is less clear (1.7 $\pm$ 1.1).

This suggests that the $rmsBoB$\ might be a better proxy in order to detect MEs.
We also observe a decrease in temperature (by a factor 3.1 $\pm$ 1.9) and the typical decreasing slope of the speed, showing the radial expansion of the ME. 
Finally, the end of the ME is clearly marked by an increase of $rmsBoB$\ and $T_{\rm p}$\ when the wake starts.

%
\section{Superposed Epochs of Different ICME Groups}
\label{sect_SE_cat_1}

When investigating the signatures of ICMEs in the {\it in situ}\ data, the difficulty in defining them comes from the variety of time profiles that ICME parameters can have. While the SEA provides a method to obtain the typical, and hence the most probable profiles of ICMEs, this implies a mixing of profiles that can strongly differ in shape. On the other hand, it is also difficult to define categories of ICMEs.
By studying whether there exist parameters that can cause differences in ICME profiles, the SEA can be used to emphasize the signature of the involved physical processes. In this section, we will use the results of the SEA so as to get the general properties of different subset of events.

\subsection{Superposed epochs for velocity-grouped ICMEs.}
\label{sect_SE_ranked_ICMEs}

\citeA{Masias-meza2016} ranked their 44 ICMEs (with a well-identified MC) by using the average absolute speed of the MCs. For example, such a ranking allows them to highlight features such as higher values of plasma parameters in the sheath and also a more asymmetric $B_{\rm tot}$ profile in the MC for fast ICMEs compared to slow ones. They also show that this velocity ranking is better than any other ranking with averaged MC properties (see their Figure 3 for the resulting ranking of other parameters). As such, in the following, we investigate how the speed can affect the profiles of the ICMEs in a similar way as \citeA{Masias-meza2016} but for the whole set of ICMEs provided in the present R\&C revisited list. This provides a statistically robust sample to test whether the results previously found apply to ICMEs with or without a well defined MC.

\begin{figure}[t!]
\centering
\includegraphics[width = 15 cm]{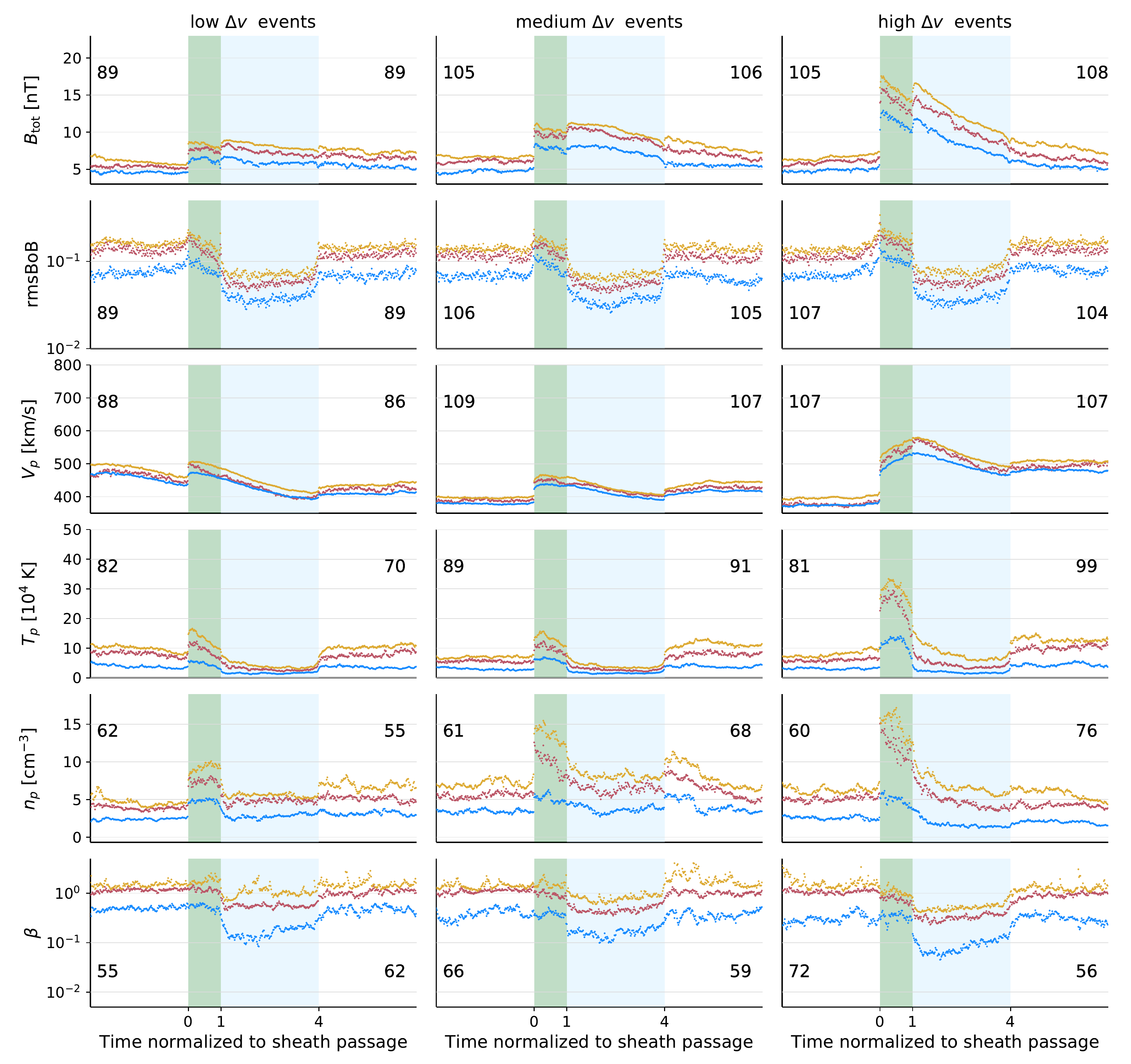}
\caption{
    Superposed epoch analysis on relatively slow, medium and relatively fast ICMEs on the left, middle and right columns, respectively. The classification in speed is performed using the difference between the ME mean speed and upstream SW mean speed.
The physical parameters shown are the same as in Figure \ref{fig_SE_sheath}.
The green area is the sheath and blue area is the ME. 
The yellow dots are the mean values, the magenta dots are the median values and the blue ones are the most probable values. The vertical scale is the same in each row. Again, the numbers on the top left (right) of each graphs are the number of events with more that 70\% of available data for the pre-SW and the sheath (ME and the wake). }
\label{fig_SE_cat_v}
\end{figure}

Since the physical interaction of the ME with the SW depends on their relative speed, in the following, we also decide to take the relative speed, rather than its absolute value, of the average ME speed compared with the average speed of the pre-SW ($V_{\rm ME}$ - $V_{\rm pre-SW}$) as a criterion to group ICMEs into three categories: relatively slow ($low$ $\Delta v$), medium ($medium$ $\Delta v$) and relatively fast ($high$ $\Delta v$). 
Note that we also investigated whether a possible correlation between $V_{\rm ME}$ - $V_{\rm pre-SW}$ and $V_{\rm ME}$ existed, and found (not shown) that there is none (some cases for which we found a negative $V_{\rm ME}$ - $V_{\rm pre-SW}$ were even found for fast events, e.g. $V_{\rm ME}$ > 500 km/s).

The frontier of the $low$ $\Delta v$\ category is set so as to select events with an ME slower than the local solar wind, i.e. with $\Delta v < 0$ km/s. By carefully studying the evolution of the sheath and ME speed profiles by binning the SEA with ICMEs with increasing speed difference,  we did not find any particular high relative speed value that could provide a frontier between the medium and relatively fast events. Thus, we set our high relative speed frontier for a $\Delta v > 55$ km/s so as to have roughly the same number of events in each group (similar statistical noise). Our group are then made of 89 $low$ $\Delta v$, 109 $medium$ $\Delta v$\ and 109 $high$ $\Delta v$. Only 307 events have enough data (less 30\% of data gaps) in order to compute $\Delta v$.

The results of the SEA applied to our three different subsets of ICMEs are shown in Figure \ref{fig_SE_cat_v}, where the same parameters as in Figure \ref{fig_SE_sheath} are shown. The left column presents the $low$ $\Delta v$\ ICMEs group, the middle one the $medium$ $\Delta v$\ ICMEs and the right one the $high$ $\Delta v$\ ICMEs. 
Some of the ICME signatures for all these three groups are similar as when all ICMEs are considered together such as a discontinuous transition between the pre-SW and the sheath, higher $B_{\rm tot}$, $T_{\rm p}$, and $n_{\rm p}$\ in the sheath, and a decrease of $rmsBoB$, $T_{\rm p}$, and $\beta$ within the ME. When comparing the categories with each other, we also find that these signatures are more pronounced for relatively fast ICMEs.

In order to quantify the values of the parameters in each substructure of ICMEs, we present in Table \ref{tab_cat_v} their most probable values averaged over the time period of each substructure, as in Table \ref{tab_all_event_wo_NS}. These values are given for the $low$ $\Delta v$\ and $high$ $\Delta v$\ ICMEs, as ranked with the relative speeds.

\begin{table}
\begin{center}
{\footnotesize
\begin{tabular}{l@{$\;\;$}c@{$\;\;$}c@{$\;\;$}c@{$\;\;$}c@{$\;\;$}c@{$\;\;$}c@{$\;\;$}c@{$\;\;$}c@{$\;$}}
\hline
 & \multicolumn{2}{c@{$\;\;$}}{pre-SW} &\multicolumn{2}{c@{$\;\;$}}{sheath} &\multicolumn{2}{c@{$\;\;$}}{ME} &\multicolumn{2}{c@{$\;$}}{post SW} \\
 & $low$ $\Delta v$ & $high$ $\Delta v$ & $low$ $\Delta v$ & $high$ $\Delta v$ & $low$ $\Delta v$ & $high$ $\Delta v$ & $low$ $\Delta v$ & $high$ $\Delta v$ \\
\hline
$B_{\rm tot}$ [nT] & 4.6 $\pm$ 0.4 & 4.9 $\pm$ 0.6 & 6.2 $\pm$ 0.9 & 11.4 $\pm$ 2.5 & 6.0 $\pm$ 1.0 & 8.6 $\pm$ 4.7 & 5.4 $\pm$ 0.8 & 5.5 $\pm$ 1.0 \\
rmsBoB [$\times 10^{-2}$] & 8 $\pm$ 3 & 7 $\pm$ 3 & 8 $\pm$ 4 & 11 $\pm$ 4 & 4 $\pm$ 2 & 4 $\pm$ 2 & 7 $\pm$ 2 & 8 $\pm$ 2 \\
$V_p$ [km/s] & 458 $\pm$ 38 $\,$ & 375 $\pm$ 8 $\,$ $\,$ & 467 $\pm$ 15 $\,$ & 507 $\pm$ 49 & 418 $\pm$ 58 $\,$ & 497 $\pm$ 67 $\,$ & 410 $\pm$ 9 $\,$ $\,$ & 480 $\pm$ 9 $\,$ $\,$ \\
$T_p$ [$\times 10^4$ K] & 3.8 $\pm$ 1.7 & 3.1 $\pm$ 0.8 & 4.7 $\pm$ 2.4 & 11.8 $\pm$ 5.6 & 1.6 $\pm$ 0.7 & 1.9 $\pm$ 1.4 & 3.6 $\pm$ 1.0 & 4.4 $\pm$ 1.6 \\
$n_p$ [cm${}^{-3}$] & 2.4 $\pm$ 0.4 & 2.6 $\pm$ 0.6 & 4.8 $\pm$ 0.7 & 4.9 $\pm$ 1.8 & 2.8 $\pm$ 0.9 & 1.7 $\pm$ 1.8 & 3.2 $\pm$ 0.8 & 2.0 $\pm$ 0.6 \\
$\beta$ & 0.5 $\pm$ 0.1 & 0.3 $\pm$ 0.2 & 0.5 $\pm$ 0.2 & 0.3 $\pm$ 0.2 & 0.2 $\pm$ 0.2 & 0.1 $\pm$ 0.1 & 0.5 $\pm$ 0.2 & 0.3 $\pm$ 0.2 \\
\hline
\end{tabular}
}
\caption{Most probable values from the distribution of the magnetic and plasma parameters, averaged over the time period of each substructure (pre-SW, sheath, ME, and wake), for relatively slow and fast categories of ICME events (see Figure \ref{fig_SE_cat_v}).
Three times the standard deviation is included after $\pm$.}
\label{tab_cat_v}
\end{center}
\end{table}

\subsubsection{Pre-ICME Solar Wind}
\label{sect_SE_Pre-ICME}

For the pre-SW, we find that the magnetic field and plasma parameters have typically similar values for the three groups.  Still, $V_{\rm p}$\ is higher for $low$ $\Delta v$\ events compared with $high$ $\Delta v$\ events.
This is a main difference with the results from \citeA{Masias-meza2016}. Indeed, in their ranking, the absolute ME speed was used, and fast events were seen to propagate in a faster SW (their Figure 4), which is not the case for our category of events as we use the relative speeds. Doing so means that some ICMEs put in the category $low$ $\Delta v$\ can therefore have a fast absolute velocity and be also propagating in a fast SW (which also explains the larger standard deviation, and the lack of correlation between the different in speeds and absolute ME speed, as discussed above). Similarly, $high$ $\Delta v$\ events tend to have slower SW at their front as this increases $\Delta v $.

\subsubsection{Sheath}
\label{sect_SE_Sheath}

Using the speed difference reinforces the contrast of the magnetic field intensity between the pre-SW and the sheath when comparing $low$ $\Delta v$\ and $high$ $\Delta v$\ events. Indeed, from Table \ref{tab_cat_v}, the ratio $r_{B} = \langle B_{sheath}\rangle / \langle B_{pre-SW}\rangle =$ 1.3 $\pm$ 0.2  and 2.3 $\pm$ 0.6 for relatively slow and fast ICMEs, respectively.
The magnetic field therefore increases much more between the pre-SW and the sheath for relatively fast events compared with relatively slow ones.  
The profile of $B$ within the sheath is also different with a steeper decrease of $B_{\rm tot}$ for $high$ $\Delta v$\ events compared with $low$ $\Delta v$\ events (Figure \ref{fig_SE_cat_v}, top panels).
A stronger $B_{\rm tot}$ in front of the sheath compared with the rear is interpreted as the consequence of the shock compression across the front shock. 

In all categories, we find that $rmsBoB$\ increases just after the front shock.
This corresponds to an increase of magnetic fluctuations after the shock.
Surprisingly, an increase of $rmsBoB$\ is also present just before the front shock, especially for the $high$ $\Delta v$\ category.  While, this could be due to an ill-located boundary, the extension of the enhancement seems too large and the shock is typically well defined.  Rather, this enhanced $rmsBoB$\ could be induced by energetic particles accelerated forward by the shock and perturbing the pre-SW magnetic field \cite<{\it e.g.,}>{blanco-cano2011,Masias-meza2016}.
This is coherent with a stronger enhancement of $rmsBoB$\ for $high$ $\Delta v$\ events as the shock is stronger and an efficient accelerator of energetic particles.

 The profile of the speed $V_{\rm p}$\ inside the sheath has a slope with an opposite sign between $low$ $\Delta v$\ and $high$ $\Delta v$\ (Figure \ref{fig_SE_cat_v}). While a front $V_{\rm p}$\ higher than the rear is generally indicative of an expansion, we observe here that relatively fast ICMEs have a lower $V_{\rm p}$\ at the front as opposed to the rear. We interpret this as a consequence of a relatively fast ME compressing the sheath more than for $low$ $\Delta v$\ events (in which a relaxation has the time to occur so that the sheath expands similarly as the ME). Such a result has not been reported by previous research, and shows the importance of studying temporal profiles along with studies of statistics of averaged parameters.

Evidences of a stronger shock and heating is also present in the temperature of protons, 
with a ratio $r_{T_p} = \langle T_{p,sheath}\rangle / \langle T_{p, pre-SW}\rangle = 1.2 \pm$ 0.8 for $low$ $\Delta v$\ events and 3.8 $\pm$ 2.1 for $high$ $\Delta v$\ between the sheath and the pre-SW. Here the large standard deviations are mainly due to a large-scale variation of the temperature (Figure \ref{fig_SE_cat_v}, fourth row).  We also observe that the separation between the values of the mean, median and most probable value is larger for $high$ $\Delta v$. This traces the $T_{\rm p}$\ distribution enlargement in the sheath. 

We also find that the density increase between the pre-SW and the sheath is not much different from the $low$ $\Delta v$\ and $high$ $\Delta v$\ events, with a ratio $r_{n_p}$ = $\langle n_{sheath}\rangle / \langle n_{p, pre-SW}\rangle = 2.0 \pm 0.4$ and $1.9 \pm 0.8$ respectively with the most probable value.  However, as for the temperature, we also observe an enlargement of the distribution for the sheath. There is indeed a higher jump for $high$ $\Delta v$\ events in density if we consider the mean and the median profile, which include more contributions of the events within the upper tail of the distribution.

The jumps found for the magnetic field and the temperature can be interpreted as the consequence of a stronger shock for $high$ $\Delta v$\ events. We also observe an increased speed difference between the sheath and the pre-SW for relatively fast ICMEs.
In contrast, the plasma $\beta$ remains about the same between the sheath and the pre-SW (Table \ref{tab_cat_v}, Figure \ref{fig_SE_cat_v}). This implies that the magnetic and plasma pressures are nearly amplified by the same factor across the shock.
   
In summary, $high$ $\Delta v$\ ICMEs create a stronger sheath. 
This is the consequence of a stronger magnetic field and plasma compression by a stronger shock, as well as by a compression by the following ME. Furthermore, the amount of overtaken SW could be larger while there is less time to evacuate the compressed plasma on the ICME sides. Then, all these processes combine to create a more heated and with a higher local magnetic field intensity sheath.  The change in density due to the relative speed is more subtle : ratios of most probable values do not show differences, however the gap between the mean and the median shows that the distribution is wider. This implies then a higher ratio of density for  $high$ $\Delta v$\ events when using the mean or the median values.

\subsubsection{Magnetic Ejecta}
\label{sect_SE_ME}

The magnetic field intensity at the ME front is higher for $high$ $\Delta v$\ events (top panel of Figure \ref{fig_SE_cat_v}). 

Relatively fast events also have a more asymmetric profile, with a more intense $B_{\rm tot}$\  at the front of the ME compared with the rear, while the $B_{\rm tot}$\ profile of relatively slow events is nearly flat.  This asymmetry causes a greater standard deviation associated with the magnetic field of the ME (Table \ref{tab_cat_v}). It shows the limitation of choosing an average value as a proxy to represent each substructure. Thus, we cannot conclude about the averaged value of $B_{\rm tot}$\ inside the ME. 
To quantify this asymmetry, we use the first moment of the magnetic field profile $c_{\rm B}$, as defined in Equation (1) of \citeA{Janvier2019}.  
A negative $c_{\rm B}$\ value means that $B_{\rm tot}$\ is higher in the first half of the profile. 
Here, we find that $c_{\rm B}$ = $-0.02$ and $-0.05$ for $low$ $\Delta v$\ and $high$ $\Delta v$\ events, respectively, which shows that the $B_{\rm tot}$\ asymmetry is stronger for $high$ $\Delta v$\ events.

Next, the fall of the $rmsBoB$\ values inside the ME is similar (0.04) among these three relative velocity categories.
These results are similar to previous analyses done with an absolute velocity ranking \cite{Masias-meza2016,Janvier2019}.

Aside from the most probable value, we can use the median and the mean value to study, for example, the extent of the distribution within the ME. 
We find that the mean and the median profile of $T_{\rm p}$\ and $n_{\rm p}$\ take more time to reach nearly constant values within relatively fast MEs.
Indeed, while the decrease is almost instantaneous after the sheath-ME boundary in the relatively slow group, it lasts almost half of the ME duration for relatively fast events. 
It means that at the beginning of the ME the distribution is wider, while it narrows further within the ME. 
We conclude that a fraction of the relatively fast events have signatures of compression in the ME front, which is in agreement with a more asymmetric $B_{\rm tot}$ profile.
One may question the accuracy of our frontiers; however, the sharp transitions at the front and the rear of the ME from the SEA, in particular for $rmsBoB$\ and $\beta$, imply that most of our MEs are well defined.

\subsubsection{Wake}
\label{sect_SE_Wake}

Finally we find that the wake speed is enhanced by a factor 1.28 $\pm$ 0.04 compared to the pre-SW speed for $high$ $\Delta v$\ events
while the wake associated to $low$ $\Delta v$\ are 0.9 $\pm$ 0.08 slower.
We see here that $high$ $\Delta v$\ events disturb the medium in which they propagate more, and also suggests that some part of the SW is slowed down behind $low$ $\Delta v$\ ICMEs.

\begin{table}[ht!]
    \centering
    \begin{tabular}{ccccc}
        \hline
        year & min SC 23 & max SC 23 & min SC 24 & max SC 24\\
        \hline
        \# events & 35 & 82 & 18 & 76 \\
        \hline
    \end{tabular}
    
    \caption{Number of events detected at 1 au within $\pm$ 2 years around 1997 (2008) for the minimum of solar cycle 23 (resp. 24) and around 2003 (resp. 2014) for the maximum of solar cycle 23 (resp. 24).}
    \label{tab_nb_events_SC}
\end{table}

\begin{figure}[t!]
\centering
\includegraphics[width = 12 cm]{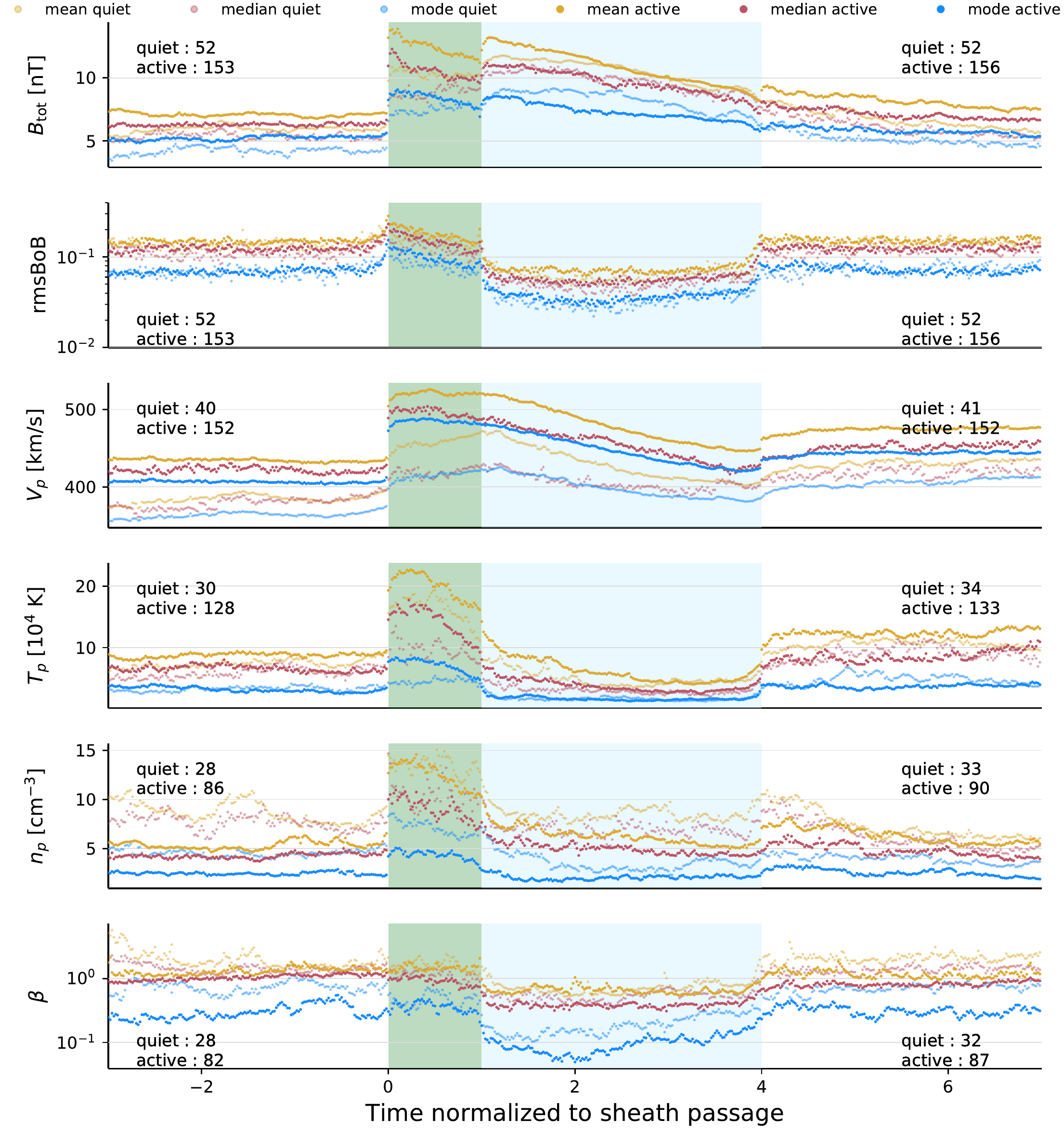}
\caption{ 
Superposed epoch analysis on events emitted during the quiet (lighter colors) and active (darker colors) period of the solar cycle. 
The physical parameters shown are the same as in Figure \ref{fig_SE_sheath}.
The green area covers the sheath while the blue area covers the ME. The yellow dots correspond to the mean values, the magenta ones to
the median value and the blue ones to the most probable values. Again, the
numbers on the top left (right) of each graphs are the number of events with
more than 70\% of available data for the pre-SW and the sheath (respectively the ME and the wake).
}
\label{fig_SE_cat_activity}
\end{figure}

\begin{table}
\begin{center}
{\footnotesize
\begin{tabular}{l@{$\;\;$}c@{$\;\;$}c@{$\;\;$}c@{$\;\;$}c@{$\;\;$}c@{$\;\;$}c@{$\;\;$}c@{$\;\;$}c@{$\;$}}
\hline
 & \multicolumn{2}{c}{pre-SW} &\multicolumn{2}{c}{sheath} &\multicolumn{2}{c}{ME} &\multicolumn{2}{c}{post SW} \\
 &  quiet & active & quiet & active & quiet & active & quiet & active \\
\hline
$B_{\rm tot}$ [nT] & 4.2 $\pm$ 0.8 & 5.2 $\pm$ 0.5 & 7.5 $\pm$ 0.9 & 8.4 $\pm$ 1.4 & 8.2 $\pm$ 2.4 & 7.3 $\pm$ 2.0 & 5.1 $\pm$ 1.4 & 5.8 $\pm$ 0.7 \\
rmsBoB [$\times 10^{-2}$] & 7 $\pm$ 3 & 7 $\pm$ 2 & 8 $\pm$ 4 & 10 $\pm$ 5 & 3 $\pm$ 2 & 4 $\pm$ 2 & 7 $\pm$ 3 & 7 $\pm$ 2 \\
$V_p$ [km/s] & 364 $\pm$ 11 $\,$ & 406 $\pm$ 4 $\,$ $\,$ &  414 $\pm$ 15$\,$ & 484 $\pm$ 9 $\,$ $\,$ & 399 $\pm$ 41 $\,$ & 447 $\pm$ 58 $\,$ & 405 $\pm$ 17 $\,$ & 443 $\pm$ 8 $\,$ $\,$ \\
$T_p$ [$\times 10^4$ K] & 3.3 $\pm$ 1.2 & 3.2 $\pm$ 1.2 & 4.5 $\pm$ 1.2 & 7.1 $\pm$ 3.2 & 1.7 $\pm$ 1.2 & 1.7 $\pm$ 1.0 & 4.8 $\pm$ 2.4 & 3.7 $\pm$ 0.9 \\
$n_p$ [cm${}^{-3}$] & 4.5 $\pm$ 1.0 & 2.4 $\pm$ 0.5 & 7.1 $\pm$ 2.2 & 4.3 $\pm$ 1.3 & 3.6 $\pm$ 2.7 & 2.1 $\pm$ 0.7 & 3.9 $\pm$ 1.3 & 2.6 $\pm$ 1.1 \\
$\beta$ & 0.7 $\pm$ 0.5 & 0.3 $\pm$ 0.2 & 0.5 $\pm$ 0.3 & 0.4 $\pm$ 0.2 & 0.2 $\pm$ 0.2 & 0.1 $\pm$ 0.1 & 0.7 $\pm$ 0.3 & 0.3 $\pm$ 0.2 \\
\hline
\end{tabular}
}
\caption{Most probable values from the distribution of the magnetic and plasma
parameters, averaged over the time period of each substructure (pre-SW, sheath,
ME, and wake), for quiet and active categories of ICME events (see Figure \ref{fig_SE_cat_activity}). Three times the standard deviation is included after $\pm$. }
\label{tab_cat_activity}
\end{center}
\end{table}

\subsection{Grouping in function of Solar Activity}
\label{sect_SE_cat_activity}

Over the 20 years of ACE observations, the Sun's activity has changed. This activity is modulated by a 11 to 13-year cycle \cite{Hathaway2010}. One may wonder whether this solar cycle has an impact on the ICME profiles observed at 1~au. Evidences collected over the years show that the solar cycle affects the occurrence of CMEs \cite<{\it e.g.,}>{Gopalswamy2003}, with more CMEs during the active period, and therefore an expected higher number of detected ICMEs. 
Previous studies have analyzed the dependence of specific properties of ICMEs and MCs ({\it e.g.,} \ size, bulk velocity or expansion) with the phase of the solar cycle \cite<{\it e.g.,}>{DassoEA12,Jian2011,Lepping2011}. Furthermore, \citeA{Hundhausen1999} found that CMEs emitted during the active period tend to be faster than the ones emitted during the quiet period. However, to our understanding there is not yet any study showing a possible dependence of the typical internal structure of ICMEs with the phase of the cycle.

Hence, we now consider a new type of grouping: we look at whether ICMEs have been emitted during the solar cycle quiet or active periods covered by ACE observations.
In our classification, an event is called ``quiet'' (respectively ``active'') if it is observed at ACE in an interval of time less than 2 years before or after a solar minimum (respectively maximum).
Since the start of the mission, ACE observed one part of the solar minimum in 1997, a full solar minimum around 2008 and two solar maxima around 2003 and 2014. 
The number of events in the R\&C~list for each extrema of the solar cycle is reported in Table \ref{tab_nb_events_SC}. 
There are around 3 to 4 times more ICMEs detected during the active phase of the solar cycle according to Table \ref{tab_nb_events_SC} with more events for the solar cycle 23 compared with solar cycle 24 in agreement with previous studies \cite<{\it e.g.,}>{Gopalswamy2003}.  
Summing each minimum and maximum we find 53 quiet events and 158 active events. 
Since the two samples have a large difference in the number of events, we verify that  selecting randomly 53 active events the SEA does not significantly change when compared with the total 158 active events.

Since the pre-SW is not affected by the ICME propagation that comes later,
these regions offer the possibility to investigate the effects of the solar
cycle on the SW itself. We compare the most probable values averaged in the pre-SW during the active and the quiet periods (Table \ref{tab_cat_activity}). 
We find that the active period SW is 1.12 $\pm$ 0.04 times faster and 1.9 $\pm$ 0.6 times less dense than its quiet period counterpart. These ratios are different than the typical ratios found when the full SW is analyzed. \cite<{\it e.g.,}>{mccomas2003,mccomas2008}.
In fact, the SW preceding ICMEs is expected to be different on average from the SW observed away from ICMEs, as follows.  ICME sources are in closed field regions of the corona, so that the SW in front of ICMEs is expected to come from the open field present nearby these closed field regions. This is especially true for the slow ICMEs which do not overtake a large fraction of SW. Then, the SW present in front of ICMEs is expected to be on average slower than away from ICMEs.

Next, we analyze the differences found in the sheath region (Figure \ref{fig_SE_cat_activity}).
The relative enhancement of the speed is similar for the pre-SW and the sheath region. The speed increases by a factor 1.19 $\pm$ 0.03 in the sheath for active events and 1.12 $\pm$ 0.04 for quiet events.
We also observe that the distribution of $B_{\rm tot}$, $T_{\rm p}$, and $n_{\rm p}$\ of active events is wider than quiet ones as shown by the separation between the most probable value and the mean (or the median) values (Figure \ref{fig_SE_cat_activity}). 
It means that during the active phase of the solar cycle, ICMEs that are emitted have a wider range of parameter values, and in particular have a sheath that can be more magnetized, hotter and denser. 

The SEA profiles of sheaths and MEs for the quiet and active periods (Figure \ref{fig_SE_cat_activity}) are similar to the SEA profiles of relatively slow and fast events, respectively (Figure \ref{fig_SE_cat_v}) especially if the means are compared. We also observe that the slope of the speed profile in the sheath is positive for quiet events while it is close to 0 (flat profile) for active events, as for $high$ $\Delta v$\ and $medium$ $\Delta v$. However, the contrast between quiet and active events is weaker than between $high$ $\Delta v$\ and $medium$ $\Delta v$\ events, so that the correlation between the relative speed and the period of activity is unclear (a further look at the relative speed ranking per solar cycle period could be done, but would significantly reduce the statistics and therefore the robustness of a conclusion).
Furthermore, $B_{\rm tot}$, $rmsBoB$,  $T_{\rm p}$, $n_{\rm p}$, and $\beta$ have similar profiles for the most probable values within the quiet and active SEAs of MEs  (Figure \ref{fig_SE_cat_activity}).  \citeA{Wu2011} found that the average magnetic field of the ME increases during the active period the solar cycle. While this difference is relatively small in our study, when comparing the
average magnetic field, this behavior disappears when using the most probable value.

We conclude that, while ICMEs are typically faster during active periods, other parameters are typically similar compared with quiet periods.  This points to a large amount of similar events in both periods.  Still, active periods are characterized by a fraction of relatively faster events which widens the ICME distributions to large parameter values. This is in agreement with \citeA{Chi2016} and \citeA{Wu2016} who also observed a larger fraction of events with more extreme speeds during the active period.

\begin{figure}[t!]
\centering
\includegraphics[width = 12 cm]{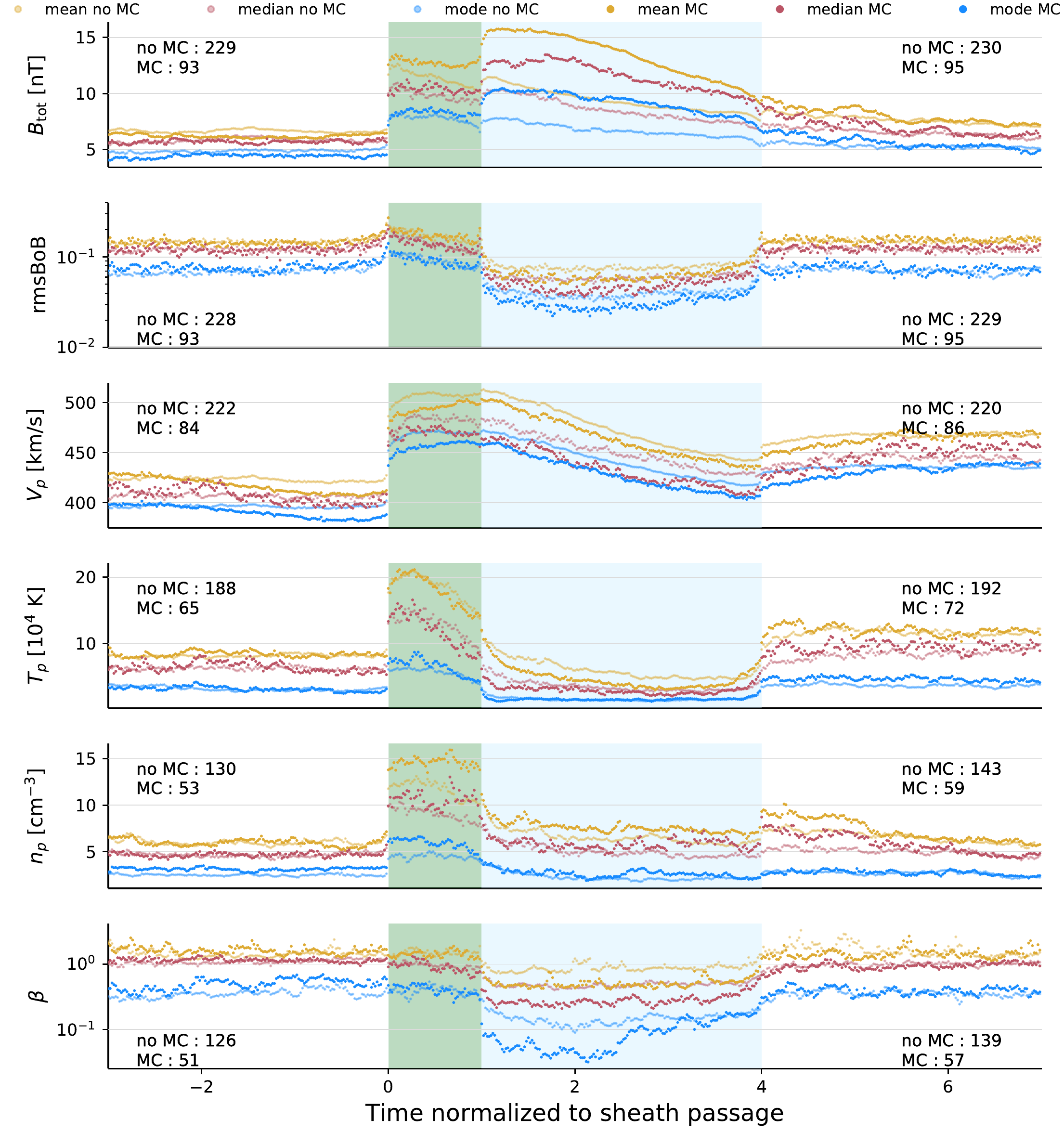}
\caption{Superposed epoch analysis on ICMEs without (lighter colors) and with (darker colors) clear MC signatures. 
The physical parameters shown are the same as in Figure \ref{fig_SE_sheath}.
The green area covers the sheath while the blue area covers the ME. The yellow dots correspond to the mean values, the magenta ones to
the median value and the blue ones to the most probable values. Again, the
numbers on the top left (right) of each graphs are the number of events with
more than 70\% of available data for the pre-SW and the sheath (respectively the ME and the wake).
}
\label{fig_SE_cat_MC}
\end{figure}

\begin{table}
\begin{center}

{\footnotesize
\begin{tabular}{l@{$\;\;$}c@{$\;\;$}c@{$\;\;$}c@{$\;\;$}c@{$\;\;$}c@{$\;\;$}c@{$\;\;$}c@{$\;\;$}c@{$\;$}}
\hline
 & \multicolumn{2}{c@{$\;\;$}}{pre-SW} &\multicolumn{2}{c@{$\;\;$}}{sheath} &\multicolumn{2}{c@{$\;\;$}}{ME} &\multicolumn{2}{c@{$\;$}}{post SW} \\
 &  no MC & MC & no MC & MC & no MC & MC & no MC & MC \\
\hline
$B_{\rm tot}$ [nT] & 4.9 $\pm$ 0.4 & 4.4 $\pm$ 0.5 & 7.8 $\pm$ 1.0 & 8.3 $\pm$ 0.9 & 6.7 $\pm$ 1.7 & 9.2 $\pm$ 2.8 & 5.3 $\pm$ 0.5 & 5.6 $\pm$ 1.6 \\
rmsBoB [$\times 10^{-2}$] & 7 $\pm$ 2 & 8 $\pm$ 3 & 9 $\pm$ 4 & 9 $\pm$ 4 & 4 $\pm$ 2 & 3 $\pm$ 2 & 7 $\pm$ 2 & 7 $\pm$ 2 \\
$V_p$ [km/s] & 396 $\pm$ 4 $\,$ $\,$ & 391 $\pm$ 18 $\,$ & 468 $\pm$ 12 $\,$ & 456 $\pm$ 15 $\,$ & 441 $\pm$ 52 $\,$ & 428 $\pm$ 52 $\,$ & 435 $\pm$ 5 $\,$ $\,$ & 431 $\pm$ 22 $\,$ \\
$T_p$ [$\times 10^4$ K] & 3.2 $\pm$ 0.9 & 3.2 $\pm$ 1.1 & 5.7 $\pm$ 2.1 & 6.5 $\pm$ 3.8 & 1.7 $\pm$ 0.9 & 1.6 $\pm$ 0.7 & 3.7 $\pm$ 0.6 & 4.7 $\pm$ 0.9 \\
$n_p$ [cm${}^{-3}$] & 2.5 $\pm$ 0.3 & 3.1 $\pm$ 0.5 & 4.5 $\pm$ 0.8 & 5.9 $\pm$ 1.7 & 2.3 $\pm$ 1.3 & 2.7 $\pm$ 1.2 & 2.7 $\pm$ 0.7 & 2.7 $\pm$ 0.7 \\
$\beta$ & 0.4 $\pm$ 0.2 & 0.5 $\pm$ 0.3 & 0.4 $\pm$ 0.2 & 0.4 $\pm$ 0.2 & 0.2 $\pm$ 0.1 & 0.1 $\pm$ 0.1 & 0.4 $\pm$ 0.1 & 0.4 $\pm$ 0.2 \\
\hline
\end{tabular}
}
\caption{
Most probable values from the distribution of the magnetic and plasma
parameters, averaged over the time period of each substructure (pre-SW, sheath,
ME, and wake), for ICMEs with or without an identified MC. Three times the standard deviation is included after $\pm$. }
\label{tab_cat_MC}
\end{center}
\end{table}

\subsection{Comparing ICMEs With and Without a MC}
\label{sect_SE_cat_MC}

In this section, we separate our sample of ICMEs into sub-samples using the detection of MC signatures. To do so, we scan by eye all the ICMEs and flag the ones that clearly show an enhanced magnetic field, a smooth rotation of one component of the magnetic field and a low proton temperature, as these are the criteria used in the literature to define MCs (see Section \ref{sect_intro}).  

In the R\&C list, the authors also added flags to each events so as to indicate whether MC signatures are observable. We ended up with 82\% of our MC ICMEs matching with their MC ICMEs (MC and MC-like flags in the R\&C list). In our sample of ICME events with a visible sheath, we only found 96 events over the 330 ICMEs that had clear MC signatures, and thus 234 events with no clear signatures. This means that around 29\% of our events have a clear MC, which is on the lower end of various estimations of the fraction of ICMEs that contain MCs \cite<>[and references therein]{Richardson2004,Chi2016}. 
In the following, we use the SEA to analyze whether profiles of ICMEs are different, in the presence or the absence of a clearly identified MC. We thus have 2 categories: events with (MC-ICMEs) and without (no MC-ICMEs) MC signatures.

We found that the physical parameters of the pre-SW have similar values for both categories (Table \ref{tab_cat_MC} and Figure \ref{fig_SE_cat_MC}) except for the pre-SW density which is weakly enhanced by a factor 1.2 $\pm$ 0.2 for MC-ICMEs compared to no MC-ICMEs. 
This similarity is expected, since the pre-SW does not hold knowledge of the magnetic structure of the ICME if there is no typical difference in their solar launch sites.
Also, for both categories, the magnetic field and plasma parameters have a clear transition between the pre-SW and the sheath (Figure \ref{fig_SE_cat_MC}).

From Table \ref{tab_cat_MC} and Figure \ref{fig_SE_cat_MC}, we find that for all the parameters except the density, the values of the most probable value as well as the width and skewness of the distributions are similar within the sheath of MC- and no MC-ICMEs. For MC-ICMEs, the density values are larger within the sheath by a ratio of 1.9 $\pm$ 0.6 compared with the averaged pre-SW. Next, the magnetic field intensity is logically larger in the MC as it is part of its definition (this was also shown in Table 1 of \citeA{Kilpua2017}). More precisely, the total magnetic field increases by a ratio $r_{B}=$ 1.4 $\pm$ 0.4 between the values found in the pre-SW and in the ME, for no MC-ICMEs, while this value is 2.1 $\pm$ 0.7 for MC-ICMEs (the standard deviations are enhanced by the large-scale variation of $B_{\rm tot}$ ). 
\citeA{Gopalswamy2006}, \citeA{Wu2011} and \citeA{Chi2016} also found that the magnetic field is higher in MC-ICMEs. They also found that the temperature is lower inside the MC compared with the ME of no MC-ICMEs. This is the case of our mean $T_{\rm p}$ profile but not for the most probable value profile.

ICME signatures are consequently more pronounced in the values of the $rmsBoB$ and $\beta$ in the ME for the MC-ICMEs, as shown by their lower values: $rmsBoB$ and $\beta$ decrease between the pre-SW and the ME, by a ratio of 2.7 $\pm$ 2.0 and 5.0 $\pm$ 5.8, respectively, for MC events, while they only decrease by a ratio 1.8 $\pm$ 1 and 2.0 $\pm$ 1.4 for no MC-events. 

These results are in agreement with the fact that we use the magnetic field as one of the major components to characterize a MC within the in situ data, while the plasma temperature, another criterion, behaves similarly in our sample between no MC-ICMEs and MC-ICMEs.
Finally, these results are coherent with the idea that no MC-ICMEs are the same as MC-ICMEs (as discussed in \citeA{Zhang2013}), but simply observed further away from the flux rope core, then with a lower field strength while plasma parameters have comparable values. 
This implies that plasma parameters are more homogeneous than the magnetic field strength within the ME, and that a large proportion of no MC-ICMEs could just as well have a MC that is not properly detected.

\section{Discussion and Conclusion}
\label{sect_conclusion}

In the present article, we detail a statistical study of the magnetic and plasma parameters within ICMEs over 20 years of ACE observations using a revisited version of the Richardson and Cane ICME catalog. This study uses the superposed epoch analysis (SEA) to extract general features from ICMEs in general but also applying it to sub-samples that split the full set into different kind of ICMEs. 

We found that for all physical parameters (such as $B_{\rm tot}$, $n_{\rm p}$, $T_{\rm p}$\ and $V_{\rm p}$), their distributions are non normal throughout the solar wind and the ICME substructures, with a long tail for large values.
Thus, we introduced the most probable value (or mode), along with the median and mean values, as proxies for the global behavior. The most probable value especially is not as sensitive to extreme values as the median or the mean values can be, while displaying them together gives a good idea of what the distribution of a physical parameter is. We see, for instance, that at the front of the ME, the temperature ($T_{\rm p}$) is typically higher, and the distribution wider, than at the rear. The front of the ME is therefore more likely to be hotter than the rear. 

Our first SEA was made with all ICMEs from our catalog that were fast enough to create a sheath. The SEA of the sheath has an increase in $B_{\rm tot}$, $V_{\rm p}$, $T_{\rm p}$\ and $n_{\rm p}$\ with a clear front discontinuity showing the sharp transition between the pre-ICME SW and this compressed plasma region. Next, the SEA on the ME emphasizes the known characteristics of this substructure: a region of smoother and enhanced magnetic field with a lower temperature as shown by the lower $rmsBoB$, $\beta$ and $T_{\rm p}$\ values.  We find that $rmsBoB$ and $\beta$ are good indicators of the presence of a ME. 

Our next SEA analyzes all the ICMEs without a sheath at the front of the ME, and presents similar features as events with a sheath with the difference of a more symmetric magnetic field profile. We interpret this result as the consequence of the absence of compression by a front sheath.  Rather, these events are typically compressed at their rear by an overtaking faster SW stream.  

Next, we ranked our ICMEs using the relative speed $\Delta v$ of the ME compared with the pre-SW since it is a key parameter for the SW-ICME interaction.
We found that the speed profile of the sheath changes from a negative slope for $low$ $\Delta v$\ events to a positive slope for $high$ $\Delta v$\ events. 
This last profile indicates a compression of the sheath by the following ME.
We however do not observe any clear increase of the magnetic field, temperature or density at the back of the sheath but rather a global enhancement in the full sheath.
This means that the ME pushing behind is able to compress the whole sheath, 
especially for $high$ $\Delta v$. This compression adds up with the front shock compression. 

On the other hand, when a $low$ $\Delta v$\ ME propagates with a velocity near the ambient SW speed, the sheath has time to adjust to the surrounding conditions (ambient SW at the front, ME at the rear), so that the sheath speed profile shows an expansion-like profile with a monotonous decrease in the speed values from the front to the rear with a slope similar to the one in the ME. Such sheath expansion is plausibly driven by the same expansion mechanism that takes places within the MC, {\it i.e.} the decrease with solar distance of the total SW pressure. 

We also found that the asymmetry of the $B(t)$ profile within the ME is larger for $high$ $\Delta v$-events, as characterized by the parameter $c_B$. This suggests that $high$ $\Delta v$\ ICMEs have a front compression of their magnetic structures causing an enhancement of the intensity of the magnetic field. We observe a much hotter sheath for $high$ $\Delta v$\ events, and these events imprint a disturbance for a longer period of time in the wake. Indeed, one ME duration is
not enough for the wake plasma to go back to the pre-ICME SW values. This means that $high$ $\Delta v$\ events are able to disturb the SW plasma during at least a day (typical duration of an ME), which might affect the propagation of another potential following ICME. 

We next investigated the effect of the solar cycle on ICME profiles.  Similarly to previous studies, we find an increased number of ICMEs during the active phase. In addition, SEA results show that there are no significant differences in the most probable values of plasma and magnetic field parameters between ICMEs emitted during the active or the quiet phases of the solar cycle. However, the distributions of these parameters within the ME are more spread to larger values during the active period. This suggests that active periods have ICMEs with characteristics common with quiet periods, together with the addition of more extreme events.

In Section \ref{sect_SE_cat_MC}, we used the observations of MC signatures as a criterion to group our ICMEs. We observed that plasma parameter profiles in the sheath or in the ME were not impacted by the presence of a MC while the $B_{\rm tot}$ profile was enhanced for events with MCs (as expected from their definition). This suggests that ICMEs with or without a detected MC are no different one from another. 
Then, one reason that only 1/3 of our sample only have a clear MC signature can be put forward as follows. When a spacecraft encounters a MC propagating in the interplanetary medium, if its trajectory crosses near the nose of the MC one would expect to find clear MC signatures.  A trajectory well away from the nose or with a crossing too far from the flux rope axis implies that at least one of the signatures (the rotation of one magnetic component) is partly lost. Furthermore, because of the interaction with the surrounding medium (sheath at the front, wake at the rear), erosion processes \cite<{\it e.g.,}>{Ruffenach2015,Dasso2006} imply that, close to the edges of MCs (front or rear), magnetic and plasma signatures of typical flux ropes 
are mixed with that of the sheath or the SW.  This implies a more difficult detection of such eroded MCs.

The results of this statistical study of ICME, in particular of different groups of ICMEs, highlight the diversity of ICME profiles observed at 1 au. For different groups of ICMEs, the observed differences and similarities allow us to probe the physical processes that happen during the propagation in the SW. Complementary studies, such as with multiple spacecraft and at different heliospheric distances, as provided by the missions Parker Solar Probe and Solar Orbiter, as well as 3D MHD simulations of CME propagation, will provide a better understanding of the early propagation processes responsible for the results found at 1 au.

\appendix

\section{Getting the most probable value}
\label{app_lognorm}

\begin{figure}[t!]
\centering
\includegraphics[width = 8 cm]{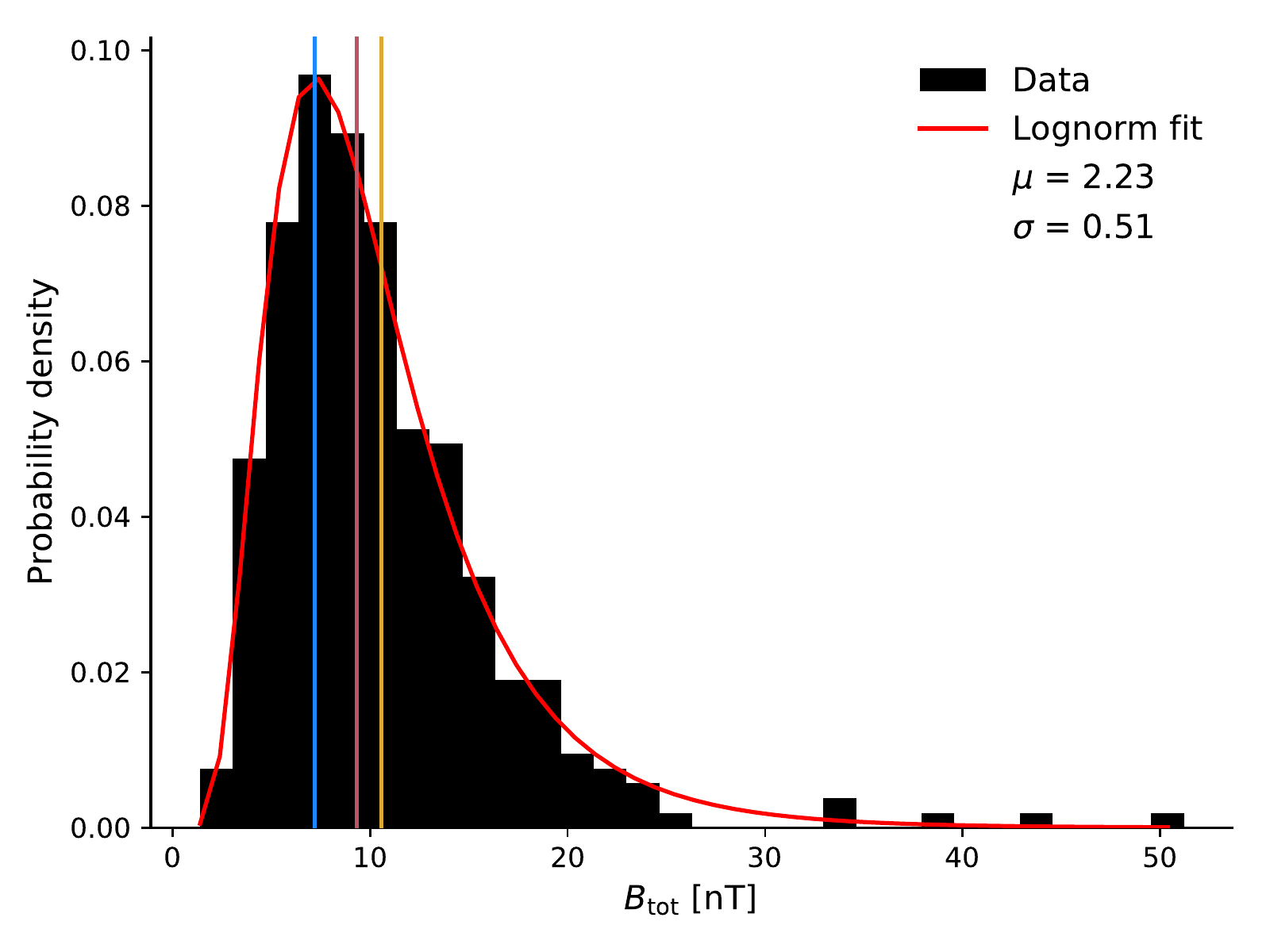}
\caption{Histogram and log-normal fit of the distribution of the magnetic field $B_{\rm tot}$\ inside the ME for all ICMEs with a sheath. The red curve is the log-normal fit and the histogram of the observed data are represented in black. The yellow line is the mean value, the magenta one the median value and the blue one is the most probable value.
The $y$-axis is the probability density. We indicate on the top right the $\mu$ and $\sigma$ parameters of the log-normal.}
\label{fig_hist}
\end{figure}

Figure \ref{fig_hist} shows the distribution (black histogram) of the magnetic field $B_{\rm tot}$\ at one time bin in the middle of the magnetic ejecta for all ICMEs with a sheath  (events of Section \ref{sect_SE_sheath}). 

Assuming a log-normal shape of the physical parameter distribution, we perform a Maximum Likelyhood Estimation (MLE) algorithm in order to get the parameters of the fitted distribution. To do so, we use the \texttt{scipy.stats.lognorm} python library (documentation at \url{https://docs.scipy.org/doc/scipy/reference/generated/scipy.stats.lognorm.html}). This allows us to estimate the parameters $\mu$ and $\sigma$, which are the mean and the standard deviation of the natural logarithm of the normally distributed variable. We can then deduce the most probable value (or mode) with the formula : mode = $e^{\mu - \sigma^2}$.

The red curve plotted on top of the histogram in Figure \ref{fig_hist} is the resulting fitted log-normal distribution. The three colored lines represent the mean value (yellow), the median value (magenta) and the most probable value (blue). We see here the shift in the position of the mean and the median value compared with the "typical" (hence most probable) value.

%
%
%
%
%
%
%
%

\acknowledgments
F.R., M.J. and F.A. acknowledge financial support from the Dim-ACAV Ile-de-France region doctoral grant as well as support from the EDOM Doctoral School at Universit\'e Paris-Saclay. 
M.J., P.D. and S.D. acknowledge financial support from the Observatoire des Sciences de l'Univers Paris-Saclay and ESTERS (Observatoire de Paris) for the visitors grant obtained.
We thank Eric Buchlin for the technical support provided in putting the ICME catalog online the IDOC/MEDOC database.

The data can be found at:
\begin{itemize}
    \item \url{http://www.srl.caltech.edu/ACE/ASC/level2/lvl2DATA_MAG.html} for the ACE magnetic {\it in situ}\ measurements.
    \item \url{http://www.srl.caltech.edu/ACE/ASC/level2/lvl2DATA_SWEPAM.html} for the ACE {\it in situ}\ plasma parameters.
    \item The list of Richardson and Cane ICMEs can be found at \url{http://www.srl.caltech.edu/ACE/ASC/DATA/level3/icmetable2.htm}
    \item The revisited catalog is at \url{https://idoc.ias.u-psud.fr/sites/idoc/files/CME_catalog/html/ACE-ICMEs-list-dates-quality-nosheath-forweb.html}
\end{itemize}


%

%

%

\end{document}